\documentclass[a4paper]{article}

\usepackage{jheppub} 
                     
\usepackage{graphicx,color}
 \usepackage{bm}
  \usepackage{bbm}
   \usepackage{amsmath}
    \usepackage{amssymb}    
     \usepackage{pifont}
\usepackage{stmaryrd}

\usepackage{braket}
\usepackage{standalone}
\usepackage{slashed}
\usepackage{multirow}
\usepackage{tabu}
\usepackage{hhline}
\usepackage{colortbl}

\newcommand{\Ds}{\displaystyle}

\newcommand{\nn}{\nonumber}

\newcommand{\Tr}{\mathrm{Tr}}
\newcommand{\sign}{\text{sign}}

\newcommand{\const}{\text{const.}}

\renewcommand{\(}{\left(}
\renewcommand{\)}{\right)}
\renewcommand{\[}{\left[}
\renewcommand{\]}{\right]}

\newcommand{\fnot}[1]{\slashed{#1}}

\title{Factorization for quasi-TMD distributions of sub-leading power}

\author[a]{Simone Rodini}
\author[b]{Alexey Vladimirov}

\affiliation[a]{CPHT, CNRS, Ecole Polytechnique, Institut Polytechnique de Paris, Route de Saclay, 91128 Palaiseau, France}
\affiliation[b]{Departamento de F\'isica Te\'orica \& IPARCOS, Universidad Complutense de Madrid, E-28040 Madrid, Spain}

\emailAdd{simone.rodini@polytechnique.edu}
\emailAdd{alexeyvl@ucm.es}

\abstract{
The quasi-transverse-momentum dependent (qTMD) distributions are equal-time correlators that can be computed within the lattice QCD approach. In the regime of large hadron's momentum, qTMD distributions are expressed in terms of standard TMD distributions via the factorization theorem. We derive the corresponding factorization theorem at the next-to-leading power (NLP), and, for the first time, we present the factorized expressions for a large class of qTMD distributions of sub-leading power. The NLP expression contains TMD distributions of twist-two, twist-three, and a new lattice-specific nonperturbative function. We point out that some of the qTMD distributions considered in this work can be employed to extract the Collins-Soper kernel using the standard techniques of different-momenta ratios. We provide NLO expressions for all the elements of the factorization theorem. Also, for the first time, we explicitly demonstrate the restoration of boost invariance of the  TMD factorization at NLP. 
}

\preprint{IPARCOS-UCM-23-025}

\begin{document} 
\allowdisplaybreaks
\maketitle

\section{Introduction}

The determination of parton distributions with lattice QCD simulations is a rapidly growing direction in the physics of strong iterations. Within the last decade, it has been raised from an abstract concept \cite{Braun:2007wv, Ji:2013dva} to practical applications (see recent reviews \cite{Cichy:2021ewm, Constantinou:2022yye}). It is foreseeable that the lattice-parton studies will soon reach a similar precision level as the experimental fits. Most importantly, lattice simulations can access parton distributions that cannot be (or are too complicated to be) measured experimentally. This is especially true for higher-twist parton distributions \cite{Bhattacharya:2021moj, Braun:2021aon, Braun:2021gvv} and transverse-momentum dependent (TMD) distributions \cite{Ebert:2018gzl, Ebert:2019tvc, Ji:2019sxk, Ji:2019ewn, Vladimirov:2020ofp}. In this work, we push the formalism of factorization theorems for lattice correlators further and derive the factorization theorem for lattice TMD distributions (usually called quasi-TMD or qTMD distributions) at sub-leading power. The derived factorization theorem connects TMD distributions of twist-three and a large set of lattice observables.

The qTMD correlator is a hadron matrix element of the form
\begin{eqnarray}
\Omega^{[\Gamma]}(y)\sim\langle P,S|\bar q(y)[\text{staple link}] \Gamma q(0)|P,S\rangle,
\end{eqnarray}
where $y$ is a space-like distance, and a staple-like gauge link connects the quark fields. The precise definition is given in sec.\ref{sec:def}. At large $P$, the qTMD correlator can be written in terms of physical TMD distributions. These relations are a particular form of the TMD factorization theorem widely used for the description of the TMD spectrum of semi-inclusive processes, for instance, \cite{Angeles-Martinez:2015sea, Scimemi:2019cmh, Bacchetta:2022awv}. In the qTMD case, the form of the factorization theorem crucially depends on the Dirac matrix $\Gamma$ that contracts the quark spinor indices. So, for $\Gamma\in\Gamma_+$ that projects both spinors to their good components, one needs the leading-power (LP) TMD factorization theorem \cite{Collins:2011zzd, Echevarria:2011epo}. This case is already well-developed theoretically \cite{Ebert:2019tvc, Ji:2019sxk, Ji:2019ewn, Vladimirov:2020ofp, Ebert:2020gxr, Ebert:2022fmh}. The next-difficulty case is $\Gamma\in\Gamma_T$ which projects a good and a bad components of quark spinor. Here one needs the next-to-leading power (NLP) TMD factorization theorem. The factorization theorem for this case is derived in this work for the first time.

From the collider-experiment view-point, the $\Gamma$-matrix is selected by the kinematic and the type of scattering process, and many polarized structures are accompanied by extra power-suppression factors making them especially difficult to access. In contrast, the lattice simulations could measure correlators with different $\Gamma$'s without conceptual complications, and the power counting of different components of $\Gamma$ is plain (i.e. $\Omega^{[\Gamma_+]}\sim P^0$ and $\Omega^{[\Gamma_T]}\sim P^{-1}$). As a matter of fact, the NLP components of qTMD correlators have been already computed. For example, within the Lorentz-invariant approaches \cite{Musch:2010ka, Musch:2011er, Engelhardt:2015xja, Yoon:2017qzo} one obtains qTMD correlator with all components of $\Gamma$ as a by-product of the computational technique. These components are usually discarded due to a lack of applicability.  One of the primary motivations for this work was to find an application for these components. 

One can expect two possible applications for qTMD distributions. They can be used to determine physical TMD distributions or to extract the Collins-Soper kernel. The latter case is especially important since it is the simplest and yet very important. In contrast to TMD distributions that parametrize partons dynamics, the Collins-Soper kernel parametrizes properties of QCD vacuum \cite{Vladimirov:2020umg}. Therefore, it can be accessed in the ratios of observables, where hadron components cancel entirely \cite{Ebert:2018gzl, BermudezMartinez:2022ctj}. Collins-Soper kernel is a universal function, and measurements from different sources can be combined together, multiplying the statistical precision \cite{Schlemmer:2021aij}. As we demonstrate in this work, the direct determination of TMD distributions from qTMD distribution of sub-leading power is not feasible (at the present moment). Nonetheless, they allow for the extraction of the Collins-Soper kernel and thus provide a new source of information for this interesting observable.

The NLP TMD factorization theorem is a relatively novel direction of research. The first steps  were made in ref.\cite{Boer:2003cm}, but the systematic development started almost twenty years later \cite{Balitsky:2017gis, Balitsky:2020jzt, Moos:2020wvd, Vladimirov:2021hdn, Ebert:2021jhy, Rodini:2022wki}. Still, many aspects of the NLP TMD factorization beyond leading perturbative order (LO) are mysterious. Thus, another main motivation for this work was to develop the factorization for the qTMD correlator at NLO till the stage of application (that is not yet done for Drell-Yan, or Semi-Inclusive Deep-Inelastic Scattering (SIDIS), although NLO expressions at the operator level are known \cite{Vladimirov:2021hdn, Rodini:2022wki}). Indeed, the expressions for the factorization theorem for the qTMD correlator are shorter but contain all principal structures. For the derivation of the factorization theorem, we use the TMD operator expansion method \cite{Vladimirov:2021hdn}, which is so far the most developed approach to NLP TMD factorization. For the first time, we explicitly demonstrate that by brining together all elements of NLP TMD factorization, one receives a valid expression satisfying all expected properties. It is not a trivial statement since the singularity structures of NLP and LP cases are different.

The paper is organized as follows. In sec.\ref{sec:def} we introduce the basic definitions and notation. In sec.\ref{sec:TMD_op_exp} we present the computation of the effective operator for the qTMD correlator in TMD factorization. Here we closely follow the method presented in ref.\cite{Vladimirov:2021hdn}. The NLO computation of coefficient functions is given in sec.\ref{sec:NLO}. The main result of this section is the bare expression for TMD factorization at LP and NLP at NLO which is presented in secs.\ref{sec:bare-position} and \ref{sec:bare-momentum} in position and momentum-fraction spaces correspondingly. The bare expression is practically useless because it contains explicit and implicit singularities and unresolved complex structures. These problems are addressed one by one in sec. \ref{sec:bare->physical}. In particular, in secs.\ref{sec:TMD-dist}, \ref{sec:Psi-distr} and \ref{sec:pole-cancel}, we provide the renormalization and evolution properties of the relevant nonperturbative functions and demonstrate the cancellation of explicit poles present in the bare version of the factorization theorem. In sec. \ref{sec:boost-inv} we discuss so-called ``special rapidity divergences'' observed in ref. \cite{Rodini:2022wki}. We explicitly demonstrate their cancellation and that this mechanism is responsible for restoring the boost invariance of TMD factorization at NLP. The complex structure of the expressions is discussed in sec. \ref{sec:complex}, in which the final form of the factorization theorem is presented. In sec. \ref{sec:practice} we discuss possible practical applications of the derived factorization formula. Specifically, we parametrize and write factorization theorems for individual measurable components of the qTMD correlator. The new and main results are collected in sec.\ref{sec:practice-at-NLP}. In appendix \ref{app:parametrization}, we summarize for convenience the used parametrizations for physical TMD distributions of twist two and three.

\section{Definition of qTMD correlator}
\label{sec:def}

In this section we introduce the main definitions and conventions.
Let us start with the defintion of the qTMD correlator:
\begin{eqnarray}
\label{def:quasiTMD}
\widetilde{\Omega}^{ij}_{q/h}(y; \mu) =Z^{-1}_W(y,L,\mu)Z^{-2}_J(\mu) \langle P,S|\bar{q}^j(y) [y; y_\perp+Lv][y_\perp+Lv;Lv][Lv;0]q^i(0)|P,S\rangle
\end{eqnarray}
where $\ket{P,S}$ is the hadron state with momentum $P$ and spin $S$  $y_\perp^\mu=y^\mu-v^\mu(vy)/v^2$, and $Z$ are the ultraviolet (UV) renormalization factors discussed below. In the convention of ref.\cite{Ebert:2022fmh}, the correlator $\Omega$ is called the qTMD beam-function. The operator in eqn. (\ref{def:quasiTMD}) is an equal-time operator, i.e. the time-components of vectors $y$ and $v$ are null. The indices $i$ and $j$ are the spinor indices of the quark fields. The notation $[a r,b r]$ identifies a straight gauge link from $b$ to $a$ in the fundamental representation of $SU(N_c)$:
\begin{eqnarray}
[a r+x,b r+x] = P\exp\( -ig\int_a^b ds \ r^\mu A_{\mu,i}(s r+x) T^i\),
\end{eqnarray}
where $r$ is any vector and $T^i$ is the generator of $SU(N_c)$ group. In eqn. (\ref{def:quasiTMD}), we do not specify the flavors of the quarks. These could be in a singlet or non-singlet combination. This choice does not modify the following computations and final results. For that reason we omit the subscript $q/h$ in the following. It is important to mention that for the singlet quantum numbers we expect that the matrix element (\ref{def:quasiTMD}) contains only the connected contribution, i.e. the contribution $\sim \langle P,S|P,S\rangle$ is subtracted.

\begin{figure}[t]
\begin{center}
\raisebox{-0.5\height}{\includegraphics[width=0.49\textwidth]{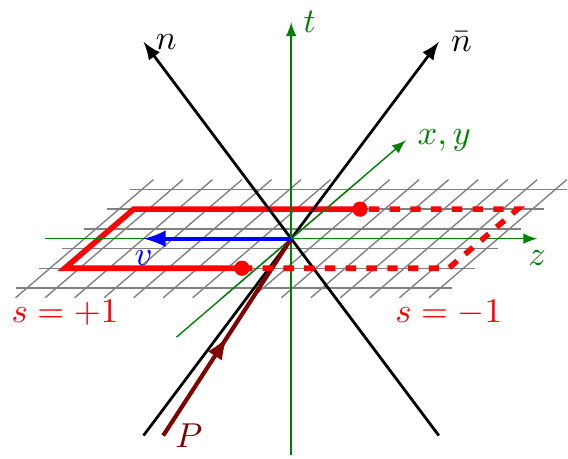}}
\raisebox{-0.5\height}{\includegraphics[width=0.49\textwidth]{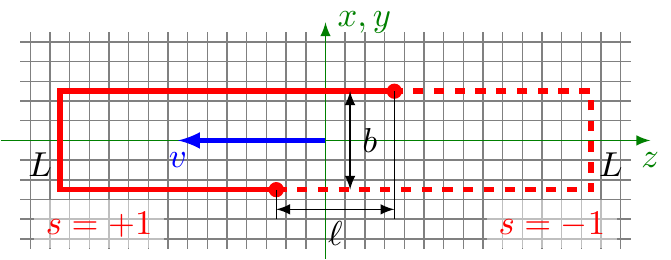}}
\end{center}
\caption{\label{fig:geometry}
Visual representation of the qTMD correlator. The left panel shows the relative definition of vectors in the space-time, with equal-time lattice at $t=0$. The right panel illustrates the geometry of qTMD operator in the spatial plane. Gray grid shows the equal-time slice of the lattice. Red lines and dots shows the gauge link and quark fields, respectively.
}
\end{figure}

The qTMD correlator is equipped with renormalization factors that make it UV finite. For simplicity, we distinguish two UV renormalizations in eqn. (\ref{def:quasiTMD}). The factor $Z_W(y,L,\mu)$ is the renormalization of UV divergences associated with the gauge links. It includes the renormalization of linear divergences \cite{Dotsenko:1979wb} and remote cusps. It could also include the scheme factors for the transition from lattice to $\overline{\text{MS}}$-scheme. The renormalization factors $Z_J$ renormalize quark fields (in axial gauge) (\ref{def:ZJ}).

Let us stress that we distinguish the notions of qTMD correlator and qTMD distribution. The qTMD correlator is the object defined in eqn.(\ref{def:quasiTMD}), and it is the outcome of the lattice computation (up to renormalization factors). The qTMD distribution is defined in the sec. \ref{sec:qTMD-distr} (eqn. (\ref{def:qTMD-distr})). It is defined such that it matches physical TMD distribution at LP/LO. From the perspective of lattice computation it involves and extra nonperturbative functions, which we refer to as $\Psi$-functions in following sections.  These functions can be identified with quasi-TMD soft factors discussed in ref.\cite{Ebert:2022fmh}.

In the limit of highly-energetic hadron and large $L$, the qTMD correlator can be expressed in terms of standard TMD distributions \cite{Ebert:2019okf, Ji:2019ewn, Vladimirov:2020ofp, Ebert:2022fmh}. In this work, we also consider the same limit. Specifically, we introduce light-cone directions $\bar n$ and $n$ identified with large and small components of the hadron's momentum, correspondingly. One has 
\begin{eqnarray}\label{def:n}
P^\mu = P^+ \bar n^\mu + \frac{M^2}{2P^+}n^\mu,
\end{eqnarray}
where $n^2=\bar n^2=0$, $(n\bar n)=1$, and $P^2=M^2$ is the mass of the hadron. The relative orientation of vectors $n$ and $\bar n$ is selected such that the vector $v$ belongs to the plane $(n,\bar n)$. Without loss of generality, we state
\begin{eqnarray}\label{def:v}
v^\mu = \frac{n^\mu-\bar{n}^\mu}{\sqrt{2}},\qquad v^2=-1.
\end{eqnarray}
The direction of the staple contour is defined by the sign of parameter $L$
\begin{eqnarray}
s=\text{sign}(L).
\end{eqnarray}
Note, that
\begin{eqnarray}
(v\cdot P)=v^-P^++\mathcal{O}\(\frac{M^2}{P^+}\)=\frac{P^+}{\sqrt{2}}+\mathcal{O}\(\frac{M^2}{P^+}\),\qquad (v\cdot P)>0,
\end{eqnarray}
is the natural large scale at play. 

The length of the gauge contour $L$ is supposed to be much larger than $y$
\begin{eqnarray}\label{L>y}
|L|\gg |y|.
\end{eqnarray}
Practically, it implies that the formulas derived in this work can be applied to the lattice measurements only after extrapolation $|L|\to\infty$. The vector $y$ is conveniently decomposed with respect to $v$
\begin{eqnarray}\label{def:y=l+b}
y^\mu= v^\mu \ell+b^\mu,
\end{eqnarray}
where the vector $b$ is entirely transverse to the scattering plane $(v,P)$, i.e., $(b\cdot v)=(b\cdot P)=0$, or equivalently, $b^+=b^-=0$\footnote{
In ref.\cite{Ebert:2022fmh}, the qTMD correlator (\ref{def:quasiTMD}) is called the beam-function in ``quasi'' scheme.  The relation between kinematic notations is the following $\eta|_\text{\cite{Ebert:2022fmh}}=L|_{\text{here}}$, $b^\mu|_\text{\cite{Ebert:2022fmh}}=y^\mu|_{\text{here}}$, $v^\mu|_\text{\cite{Ebert:2022fmh}}=v^\mu|_{\text{here}}$, $\delta^\mu|_\text{\cite{Ebert:2022fmh}}=(0,0,0,\ell)|_{\text{here}}$, $(0,b_T^x,b_T^y,0)|_\text{\cite{Ebert:2022fmh}}=b^\mu|_{\text{here}}$, and $\tilde b^z|_\text{\cite{Ebert:2022fmh}}=\ell|_{\text{here}}$.
}. In eqn.(\ref{def:quasiTMD}) $y_\perp^\mu=b^\mu$. For the explicit realisation of vectors see sec.\ref{sec:qTMD-distr}.

The general structure of the qTMD correlator resembles those of hadron tensors for Drell-Yan or semi-inclusive deep-inelastic scattering (SIDIS) processes. To  complete the analogy, we introduce the current
\begin{eqnarray}
J^i_v(y;L) = [\infty b+L v,y+L v][y+L v,y] q^i(y),
\end{eqnarray}
which transforms as a fundamental representation of $SU(N_c)$. Using this notation, the quasi-TMD reads
\begin{eqnarray}
\label{def:quasiTMD-JJ}
\widetilde{\Omega}^{ij}(y; v, P, S) = \langle P,S|\bar J_v^j(y) J_v^i(0)|P,S\rangle,
\end{eqnarray}
where $\bar J_v^\mu=(J_v^\mu)^\dagger \gamma^0$. We also introduce the notation
\begin{eqnarray}
\label{def:quasiTMD-JJ-Gamma}
\widetilde{\Omega}^{[\Gamma]} =\frac{1}{2}\Tr\(\Gamma \widetilde{\Omega}\)= \langle P,S|\bar J_v(y)\,\frac{\Gamma}{2}\, J_v(0)|P,S\rangle,
\end{eqnarray}
where $\Gamma$ is a Dirac matrix, and we have suppressed the arguments of qTMD. 

Along the paper, we operate both in the position and in the momentum-fraction representations. Both representations have specific advantages and disadvantages. We strictly follow the convention to decorate functions in position space by a tilde. The relation between qTMD correlators in position and momentum-fraction spaces is
\begin{eqnarray}\label{def:Fourier-OMEGA}
\Omega^{[\Gamma]}(x,b)=\int_{-\infty}^{\infty} \frac{d\ell}{2\pi} 
e^{-ix\ell P^+} \widetilde{\Omega}^{[\Gamma]}(\ell,b).
\end{eqnarray}
Note, that the integration over $\ell$ violates the condition (\ref{L>y}). We understand the transformation (\ref{def:Fourier-OMEGA}) formally and restrict $x\gg (LP_+)^{-1}$.

\section{TMD operator expansion for qTMD operator}
\label{sec:TMD_op_exp}

In this section, we derive the bare form of the factorization theorem for the qTMD correlator using the method of TMD operator expansion, and compute the (bare) coefficient functions at NLO. The derivation follows the one for the correlators of electro-magnetic currents presented  in details in ref.\cite{Vladimirov:2021hdn}. Therefore, we skip most of the conceptual discussion and point out only the specific features of the qTMD case. The main result from the calculations presented in this section is the bare form of factorization theorem given in secs. \ref{sec:bare-position} and \ref{sec:bare-momentum}.

\subsection{Effective operator and field counting}
\label{sec:counting}

The procedure of the TMD operator expansion starts with the functional-integral formulation of the qTMD correlator. It can be easily done, since the operator in (\ref{def:quasiTMD}) is an equal-time operator. We have
\begin{eqnarray}\label{def:func0}
\widetilde{\Omega}_{\text{bare}}^{[\Gamma]}(y)&=& \int [D\bar q D q D A] e^{iS_{\text{QCD}}}\phi^*(P,S)\bar J_v(y) \frac{\Gamma}{2}J_v(0) \phi(P,S),
\end{eqnarray}
where $\phi$ is the hadron's wave function, and $S_{\text{QCD}}$ is the QCD action. 

Next, we declare the parton model for the hadrons. The parton model consists in the statement that the constituent fields of a fast hadron are almost free and that their traverse momentum is suppressed in comparison to their longitudinal momentum. In other words, the hadron consists of collinear fields, which we label by subscript $\bar n$. They obey the counting
\begin{eqnarray}\label{count:nbar}
\{\partial_+, \partial_-, \partial_T\}q_{\bar n} & \lesssim & P_+\{1, \lambda^2, \lambda\}q_{\bar n},
\\\nn
\{\partial_+, \partial_-, \partial_T\}A^\mu_{\bar n} & \lesssim & P_+\{1, \lambda^2, \lambda\}A^\mu_{\bar n},
\end{eqnarray}
where $\lambda\sim M/P^+$ is a small parameter. The sign $\lesssim$ indicates that the partons' momenta are not restricted from below and thus includes also lower-counting modes. The momentum counting rules and the QCD equation of motions (EOMs) fix the counting for the components of the fields. It is straightforward to demonstrate that
\begin{eqnarray}\label{count:q}
&&\{\xi_{\bar n},\eta_{\bar n}\}=\Big\{\frac{\gamma^-\gamma^+}{2}q_{\bar n},\frac{\gamma^+\gamma^-}{2}q_{\bar n}\Big\}\sim \{\lambda,\lambda^2\},
\\\label{count:A}
&&\{A_{\bar n+},A_{\bar n-},A_{\bar n T}\}\sim \{1,\lambda^2,\lambda\}.
\end{eqnarray}

The main difference of the TMD factorization from collinear factorization is the counting rule for the distance $y$. It has a large transverse component $b\sim (\lambda P_+)^{-1}$. As a consequence, the transverse derivatives of collinear field accompanied by $b$ have a unity counting, $b^\mu\partial_\mu q_{\bar n}\sim 1$, which result in the TMD-type of operators. Simultaneously, we should guarantee that the anti-collinear momentum of partons remains suppressed, $y^+ \partial_-q_{\bar n}\sim \lambda$ (at least), and that collinear derivatives are not magnified, i.e. $y^- \partial_+q_{\bar n}\sim 1$ (at most). The latter is required in order to keep away of the small-$x$ effects. In our definition $y^+\sim y^-\sim \ell$. These requirements uniquely fix the scaling
\begin{eqnarray}\label{count:y}
\{\ell,b\}\sim P_+^{-1}\{1,\lambda^{-1}\}.
\end{eqnarray}
Note that these constraints imply that $\ell\ll b$, which should be fulfilled in lattice simulations.

\begin{figure}
\begin{center}
\includegraphics[width=0.3\textwidth]{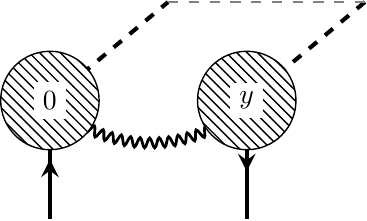}
\end{center}
\caption{\label{fig:exchange-diagram} Example of the diagram topology that has suppressed counting. In this diagram the size of propagator is $\sim b^2$, and thus it is $\sim b^{-2}\sim \mathcal{O}(\lambda^2)$ at the best.}
\end{figure}

The counting (\ref{count:y}) does not fully incorporate the nonperturbative components of eqn.(\ref{def:func0}). The extra source of nonperturbative corrections are the gluon fields present in the Wilson line. These fields can couple between Wilson links separated by large-$b$. Such interactions are nonperturbative and to extract their contribution, we introduce the $v$-collinear modes with the counting
\begin{eqnarray}\label{count:v}
\{\partial_+, \partial_-, \partial_T\}q_{v} & \lesssim & P_+\{\lambda^2, \lambda^2, \lambda\}q_{v},
\\\nn
\{\partial_+, \partial_-, \partial_T\}A^\mu_{v} & \lesssim & P_+\{\lambda^2, \lambda^2, \lambda\}A^\mu_{v}.
\end{eqnarray}
There is no necessity to split the fields further and introduce extra modes, because collinear and $v$-collinear modes accumulate all nonperturbative effects. 

To integrate the perturbative component, we split the fields in the functional integral as
\begin{eqnarray}
q=\psi+q_{\bar n}+q_v,\qquad A^\mu=B^\mu+A^\mu_{\bar n}+A^\mu_v,
\end{eqnarray}
where $\psi$ and $B$ are the dynamical fields, which do not satisfy the counting (\ref{count:nbar}, \ref{count:v}). This decomposition covers all configurations within the functional integral. There is, however, a double counting, which takes place if the (anti)collinear momentum of the collinear field became too soft, $\{\partial_+, \partial_-, \partial_T\}q\sim \{\lambda^2, \lambda^2, \lambda\}q$.

There are two popular approaches to resolve the issue of double-counting. The first is to introduce an extra cutting rule in the overlap region, which eliminates the overlap on the level of the functional integral. This approach is used in refs. \cite{Balitsky:2017gis, Balitsky:2020jzt}. The second approach is to assume that hadrons  do no contain the soft components (which is valid for the non-small-x approximation), and subtract double-counting contribution by division of the functional integral by corresponding vacuum contribution \cite{Manohar:2006nz, Collins:2011zzd}. This contribution is called the soft factor and  denoted as $S(y)$. In this work, we utilize the second approach. The resulting functional integral reads
\begin{equation}\label{def:func0}
\widetilde{\Omega}_{\text{bare}}^{[\Gamma]}(y)= \!\int 
[D\bar q_{\bar n} D q_{\bar n} D A_{\bar n}]
[D\bar q_{v} D q_{v} D A_{v}]
 e^{iS_{\text{QCD}}[q_{\bar n},A_{\bar n}]+iS_{\text{QCD}}[q_{v},A_v]}\phi^*(P,S)\frac{\mathcal{W}^{[\Gamma]}_{eff}(y)}{S(y)} \phi(P,S),
\end{equation}
where
\begin{eqnarray}
\mathcal{W}^{[\Gamma]}_{eff}(y)=
\int [D\bar \psi D \psi DB] e^{iS_{\text{int.}}}\(\bar J_v(y) \frac{\Gamma}{2}J_v(0)\)[\psi+q_{\bar n}+q_v,B+A_{\bar n}+A_v].
\end{eqnarray}
Here, $S_{\text{int.}}=S_{\text{QCD}}[\psi+q_{\bar n}+q_v,B+A_{\bar n}+A_v]-S_{\text{QCD}}[q_{\bar n},A_{\bar n}]-S_{\text{QCD}}[q_{v},A_v]$ is the background-field Lagrangian with two background fields. Explicit expression for $S_{\text{int.}}$ can be found in appendix A of ref.\cite{Vladimirov:2021hdn}. 

The definition of the soft factor depends entirely on the shape of the overlap region, which is defined by the counting rules. Since in the present case the overlap region coincides with the case for the ordinary TMD factorization theorem \cite{Collins:2011zzd, Echevarria:2011epo}, the soft factor is the usual TMD soft factor. 

In the form (\ref{def:func0}) the factorization theorem does not require a proof. In the sense  that the expression (\ref{def:func0}) is already factorized. Indeed, the effective operator $\mathcal{W}^{[\Gamma]}_{eff}$ is a polynomial in background fields, and all interaction structure is already collected into action exponents. The signal of the factorization violation would be a mismatch of the singularity structures between poles of nonperturbative elements, coefficient functions, and $S(y)$. As we demonstrate in sec.\ref{sec:bare->physical}, in the present case all singularities cancel in-between terms. This confirms the factorization statement.

One of the advantages of the background-field method is the possibility of fixing different types of gauges for the dynamical and for each background sector. We use the standard choice of background gauge for the dynamical gluon \cite{Abbott:1980hw}. For the background fields, we use the light-cone gauges, because this choice essentially simplifies the computation. We define the gauge-fixing conditions
\begin{eqnarray}\label{def:gauge}
A_{\bar n}^+=0,\qquad A_v^-=0.
\end{eqnarray}
The light-cone gauge is to be supplemented by the appropriate boundary conditions for the transverse component of the fields. We set
\begin{eqnarray}\label{def:boundary}
\lim_{z^-\to s\infty}A^{\mu_{T}}_{\bar n}(z)=0,\qquad
\lim_{z^+\to -\infty}A^{\mu_T}_{v}(z)=0.
\end{eqnarray}
This choice follows from the analysis of the integrals at one-loop, which are presented below. Boundary conditions (\ref{def:boundary}) are fixed to nullify the gluon interaction at spatial infinity, which corresponds to the diagrams with interaction with transverse links. For a more detailed discussion we refer to sec.3 of ref.\cite{Vladimirov:2021hdn}. As consequence of (\ref{def:gauge}) and (\ref{def:boundary}), the components of the gluon field can be expressed via the field-strength tensor
\begin{eqnarray}\label{A->F}
A_{\bar n}^\mu(z)=-\int_{s\infty}^0 ds F^{\mu+}(s n+z),
\qquad
A_{v}^\mu(z)=-\int_{-\infty}^0 ds F^{\mu-}(s \bar n+z).
\end{eqnarray}

Let us note that the gauge-fixation condition for $v$-collinear field (\ref{def:gauge}) is somewhat redundant. Indeed, the counting rules for the components $A_v$-field states that $A_{v}^- \sim A_{v}^+ \sim \lambda^2$, and thus the sensitivity to $A_v^-$ is beyond our accuracy. In this case, one cannot justify the choice of gauge-fixing solely by counting arguments (as it could be done for factorization of cross-section). To fix it, one needs to perform one-loop computation and restore the Wilson line from the gluon interaction. Such computation was done in ref.\cite{Vladimirov:2020ofp}, and conditions (\ref{def:gauge}, \ref{def:boundary}) corresponds to it.

The integral for the effective operator $\mathcal{W}$ is to be taken by means of the perturbative expansion. Throughout this process, the background fields are considered as external classical fields, which satisfy the QCD equation of motions (EOMs). The loop-coordinates have an effective counting $\sim 1/P^+$ for all components, despite the loop integrals span the whole space. Therefore, the power-unsuppressed interactions are effectively confined in small volumes around currents at $0$ and $y$. The diagrams that include an exchange between these volumes (see fig. \ref{fig:exchange-diagram}), contain propagators in the distance $(x-y)$ (with $x$ being a loop-coordinate). Such propagators lead to a suppression factor $\sim b^{-2}\sim \lambda^2$. In other words, such diagrams are NNLP at least. The more propagators connect the volumes the higher is the suppression. The main conclusion of this hierarchy is that LP and NLP contributions to the effective operator come from the diagrams without exchanges between $0$ and $y$ positions. The whole set of this diagrams can be presented as the product
\begin{eqnarray}\label{W=JJ}
\mathcal{W}^{[\Gamma]}_{eff}(y)=\bar{\mathcal{J}}(y)\frac{\Gamma}{2}\mathcal{J}(0)+\mathcal{O}(\lambda^2),
\end{eqnarray}
where
\begin{eqnarray}
\label{eq:fund-current}
\mathcal{J}(y)=
\int [D\bar \psi D \psi DB] e^{iS_{\text{int.}}}J_v(y)[\psi+q_{\bar n}+q_v,B+A_{\bar n}+A_v].
\end{eqnarray}
This relation is straightforward to proof using the combinatorial formula for the disconnected diagrams. Thus, to receive LP and NLP expressions one needs to derive the NLP expansion for the effective current $\mathcal{J}$ only.

Note, that in eqn.(\ref{W=JJ}) we indicated the order of correction $\mathcal{O}(\lambda^2)$ relatively to the leading term. We will do the same in all subsequent sections.

\subsection{Effective current at LO}

\begin{figure}
\begin{center}
\includegraphics[width=0.5\textwidth]{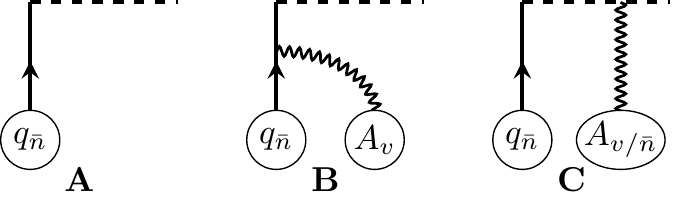}
\end{center}
\caption{\label{fig:LO-diagram} Diagrams contributing to the LO expression for the effective current at LP and NLP. The dashed line indicates the Wilson line. The labels in blobs indicate the type of external field. The diagram \textbf{C} vanishes in our choice of the gauge.}
\end{figure}

The LP and tree order of the  effective current $\mathcal{J}$ is given by diagram \textbf{A} in fig.\ref{fig:LO-diagram}. It reads
\begin{eqnarray}
\mathcal{J}_{\mathbf{A}}(0)&=&P\exp\[-ig\int_L^0 ds v^\mu(A_{v,\mu}(s v)+A_{\bar n,\mu}(s v))\]
q_{\bar n}(0).
\end{eqnarray}
Here, we observe that the Wilson line contains only the fields which are either zero due to the gauge choice (\ref{def:gauge}), or suppressed as $\mathcal{O}(\lambda^2)$. As the result, we obtain $\mathcal{J}_{\mathbf{A}}(0)=q_{\bar n}(0)+\mathcal{O}(\lambda^2)$, i.e. there are no $v$-collinear fields. If expanded further, the final formula would contain the uncompensated rapidity divergences, in the $v$-collinear sector, which indicates the missed contribution. The more formal consideration has been done in ref.\cite{Vladimirov:2020ofp}, where it has been shown at one loop that the $v$-collinear sector is represented by the Wilson line $[Lv,0]$. The missed source of the enhancement is the integral in the Wilson line $\int_L^0 ds\sim L$. 

To formalize this observation we introduce
\begin{eqnarray}\label{def:H}
H^\dagger(z)=P\exp\[-ig\int_L^0 ds \,v^\mu A_{v,\mu}(s v+z)\]\sim \mathcal{O}(\lambda^0).
\end{eqnarray}
As we demonstrate later, such assumption leads to the correct factorization theorem at LP and NLP, i.e. we observe the cancellation of divergences and expected properties of the factorization.

Let us also note the possibility to have a contribution $\sim q_v$. Generally speaking, such term is also of LP or NLP. However, within the matrix element it could be coupled only to a similar term $\sim \bar q_v$ in $\bar{\mathcal{J}}$, due to the fermion number conservation. Consequently, the outcome of such term is $\sim \langle P,S|P,S\rangle \langle 0|\bar q_v q_v |0\rangle$, i.e. it is disconnected. We drop disconnected contributions according to our initial assumption. The NLP terms $\sim q_v$ also do not contribute to the connected part. The first non-zero contribution of such type appears only at N$^2$LP.

In this way, the LP/LO expression to the effective current is
\begin{eqnarray}\label{LP:LO}
\mathcal{J}_{\text{LP/LO}}(z)=H^\dagger(z)\xi_{\bar n}(z),
\end{eqnarray}
where $\xi_{\bar n}$ is the good component of the quark field defined in eqn.(\ref{count:q}). 

To receive the NLP terms one should consider diagrams shown in fig.\ref{fig:LO-diagram}. The computation yields
\begin{eqnarray}\label{NLP:LO-inter}
\mathcal{J}_{\text{\textbf{A}/NLP}}(z)=H^\dagger(z)\eta_{\bar n}(z),\qquad
\mathcal{J}_{\text{\textbf{B}/NLP}}(z)=\frac{ig}{2}H^\dagger(z)\gamma^+\fnot A_{v T}(z)\frac{1}{\partial_+}\xi_{\bar n}(z),
\end{eqnarray}
where $g$ is the QCD coupling constant, and the inverse derivative is defined as
\begin{eqnarray}
\frac{1}{\partial_+}f(x)=\int_{s\infty}^0dz^- f(x+z^-n).
\end{eqnarray}
The diagram \textbf{C} vanishes due to our choice of the gauge conditions. 

The form of expressions (\ref{NLP:LO-inter}) is not unique but could be modified using EOMs. This presents the problem of fixing the operator basis for the sub-leading power computation. It is known that, generally, sub-leading power distributions mix with the leading power distributions under the renormalization. Therefore, the best choice for the basis is the one that nullifies the mixing. This is accomplished by sorting the operators with respect to the twist, that is the Lorenz-invariant characteristic of operator. The twist of operator is computed by the usual dimension-minus-spin rule, where spin is projected to the collinear direction. Distribution with different twists do not mix. In the present case the twist-decomposition must be done for collinear operator only, because the $v$-collinear operators already have the minimal twist at NLP. 

The operator $\xi_{\bar n}$ is of twist-one, and it cannot be reduced. The twist of  operator $\eta_{\bar n}$ (that appear in $\mathcal{J}_{\text{\textbf{A}/NLP}})$ is not defined. Applying EOMs in the massless quark approximation, the field $\eta_{\bar n}$ can be expressed via $\partial_\mu \xi_{\bar n}$ (the total derivative of twist-one operator) and $A_{\bar n,\mu}\xi_{\bar n}$ (the twist-two operator). These operators have independent renormalizations, and thus represent our go-to choice for the basis. 

Combining together the expressions for diagrams and applying EOMs, we obtain the effective current in the simple form
\begin{eqnarray}\label{NLP:LO}
\mathcal{J}_{\text{NLP/LO}}(z)=-\frac{1}{2}H^\dagger(z)\frac{\gamma^+}{\partial_+}\(\fnot \partial_T-ig \fnot A_{\bar n T} -ig \fnot A_{v T}\)\xi_{\bar n}(z),
\end{eqnarray}
where $\partial_+$ acts only to the $\bar{n}$-collinear fields. The expression for the conjugated current reads
\begin{eqnarray}
\bar{\mathcal{J}}_{\text{NLP/LO}}(z)=-\frac{1}{2}\bar \xi_{n}(z) \(\overleftarrow{\fnot \partial_T}+ig \fnot A_{\bar n T} +ig \fnot A_{v T}\)\frac{\gamma^+}{\overleftarrow{\partial_+}}H(z).
\end{eqnarray}

\subsection{Effective current at NLO}
\label{sec:NLO}

\begin{figure}
\begin{center}
\includegraphics[width=0.85\textwidth]{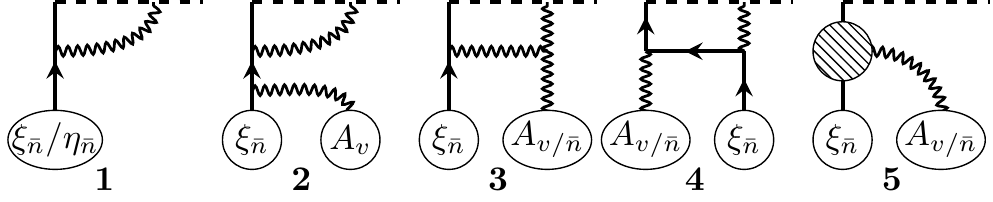}
\end{center}
\caption{\label{fig:NLO-diagram}  Diagrams contributing to the NLO expression for the effective current at LP and NLP. The dashed line indicates the Wilson line. The labels in blobs indicate the type of external field. The diagram \textbf{5} represents all possible one-loop extensions of the quark-gluon vertex.}
\end{figure}

At NLO, the expression for the effective current acquires the coefficient functions. The expressions (\ref{LP:LO}) and (\ref{NLP:LO}) take the form
\begin{eqnarray}\label{def:J+coef}
\mathcal{J}(z)&=&
H^\dagger(z)\widehat{C}_1\xi_{\bar n}(z)
-\frac{1}{2}
\gamma^+\gamma^\mu H^\dagger(z)\widehat{C}_1 \frac{\partial_\mu}{\partial_+}\xi_{\bar n}(z)
\\\nn &&
+\frac{ig}{2}
\gamma^+\gamma^\mu H^\dagger(z)\frac{1}{\partial_+}\widehat{C}_2 A_{\bar n \mu}(z) \xi_{\bar n}(z)
+\frac{ig}{2}
\gamma^+\gamma^\mu H^\dagger(z)A_{v \mu}(z)\frac{1}{\partial_+}\widehat{C}_{2v} \xi_{\bar n}(z)
+\mathcal{O}(\lambda^2).
\end{eqnarray}
The coefficient functions $\widehat{C}_i$ are integral operators that act on the collinear fields. The $v$-collinear fields do not participate in the integral convolution. The factors in eqn. (\ref{def:J+coef}) are normalized such that $\widehat{C}_i=1+\mathcal{O}(a_s)$.

The diagrams contributing to the NLO coefficient functions are shown in fig.\ref{fig:NLO-diagram}. The technique of calculation is presented in details in ref.\cite{Vladimirov:2021hdn}. 

Computing the diagram \textbf{1} with the external field $\xi_{\bar n}$ we receive the bare coefficient function for LP operator. It reads
\begin{eqnarray}\label{NLO:C1-position}
\widehat{C}_1\xi_{\bar n}(0)=
\xi_{\bar n}(0)-2a_sC_F\frac{1-\epsilon}{1-2\epsilon}\Gamma(-\epsilon)\(\frac{-v^2}{4}\)^\epsilon \int_{s\infty}^0 \frac{d\sigma}{\sigma (\sigma^2)^{-\epsilon}} \xi(v^-\sigma)+\mathcal{O}(a_s^2),
\end{eqnarray}
where $\epsilon$ is the parameter of dimensional regularization $d=4-2\epsilon$, $a_s=g^2/(4\pi)^{d/2}$, and $C_F=(N_c^2-1)/2N_c$ for $SU(N_c)$ group. We stress that the order of powers in the denominator is $(\sigma^2)^{-\epsilon}$ ($\neq \sigma^{-2\epsilon}$), which is important to produce the correct complex part of the expression in momentum-fraction space.

The coefficient function for the operator $\partial_\mu \xi_{\bar n}$ is also computed from the diagram $\mathbf{1}$. It is collected from two distinct parts. The first part is the diagram $\mathbf{1}$ with the external field $\xi_{\bar n}$ computed up to a transverse derivative (NLP contribution). The second part is the diagram $\mathbf{1}$ with the external field $\eta_{\bar n}$ at LP, which after application of EOMs contains a term proportional to $\partial_\mu \xi_{\bar n}$.

The coefficient function $\widehat{C}_2$ is obtained from the diagrams \textbf{1} (after application of EOM to $\eta_{\bar n}$), \textbf{3} and \textbf{4}. The diagrams of type \textbf{5} are zero in the dimensional regularization, due to the absence of a Lorentz-invariant scale parameter in the loop-integral for the single propagator. The NLO expression for $\widehat{C}_2$ reads
\begin{eqnarray}\label{NLO:C2-position}
\widehat{C}_2f(0,0)&=&
f(0,0)+2a_s\Gamma(-\epsilon)\(\frac{-v^2}{4}\)^\epsilon \int_{s\infty}^0 \frac{d\sigma}{\sigma (\sigma^2)^{-\epsilon}}\int_0^1 d\alpha \Big[
\\\nn &&
C_F \frac{1-3\epsilon }{1-2\epsilon}f(\sigma,\sigma)
-\(C_F-\frac{C_A}{2}\)\(f(\alpha \sigma,\sigma)+\frac{f(0,\sigma)}{1-2\epsilon}\)
\\\nn &&
+\frac{C_A}{2}\(f(\sigma,\alpha\sigma)-\frac{2\epsilon}{1-2\epsilon}f(\sigma,0)\)\Big]
 +\mathcal{O}(a_s^2),
\end{eqnarray}
where $C_A=N_c$ and we use the convenient notation
$$
f(x,y)=A_{\bar n\mu}(v^-x)\xi_{\bar n}(v^-y).
$$

The coefficient function $\widehat{C}_{2v}$ is obtained from the diagrams \textbf{2}, \textbf{3}, and \textbf{4}.  The diagrams of type \textbf{5} are zero in the dimensional regularization, similarly to the case of $\widehat{C}_2$. Note that the absence of contribution from diagrams $\mathbf{5}$ happens solely due to counting rules for the field $A_{v}$ (\ref{count:v}), and usually such diagrams contribute to the coefficient function (see, e.g., the case of TMD factorization for Drell-Yan process \cite{Vladimirov:2021hdn}). As a result of computation, one finds that the NLO expression for $\widehat{C}_{2v}$ is equal to (\ref{NLO:C1-position}). At the moment we cannot provide a solid argument  that this equality is preserved beyond NLO. Therefore, we conservatively state
\begin{eqnarray}\label{NLO:C2v-position}
\widehat{C}_{2v}=\widehat{C}_{1}+\mathcal{O}(a_s^2).
\end{eqnarray}

It is interesting to observe that a similar relation holds between coefficient functions in the ordinary TMD factorization theorem: the regularized (with $\epsilon<0$) NLP coefficient function (see (6.13) in \cite{Vladimirov:2021hdn}) at vanishing gluon momentum coincides with regularized LP coefficient function (see (6.12) in \cite{Vladimirov:2021hdn}), i.e. $C^{\text{\cite{Vladimirov:2021hdn}}}_2(x_2=0,\epsilon)=C^{\text{\cite{Vladimirov:2021hdn}}}_1(\epsilon)$. We note that such relation does not hold if expansion over $\epsilon$ is taken before $x_2\to0$ limit (compare (6.14) and (6.15) in \cite{Vladimirov:2021hdn}).

We do not know the fundamental reason for a relation between NLP and LP coefficient functions in TMD factorization. The most plausible explanation is that this is a particular case of Ward identities at vanishing momentum, also known as soft-gluon theorems \cite{Bern:2014vva, Hamada:2018vrw, Li:2018gnc}. If so, the relations must hold at all perturbative orders, and could serve as an additional demonstration of the correctness of our computation.

\subsection{Bare qTMD correlator at NLP (position space)}
\label{sec:bare-position}

We now combine the expressions for the effective currents to form the factorization theorem for the effective operator (\ref{W=JJ}). The LP term is given by the product of LP currents (\ref{LP:LO}), whereas the NLP terms are the product of one LP and one NLP currents. 

Let us schematically summarize the necessary steps to obtain the desired factorized expression: \textit{i)} multiply effective currents into the effective operator (\ref{W=JJ}); \textit{ii)} write expressions in a generic gauge by assigning the light-cone links and rewriting gluon fields via the gluon field-strength tensors (\ref{A->F}); \textit{iii)} recouple the color indices such that products of collinear and $v$-collinear fields are independently color-neutral. Finally, the hadron matrix element is taken, and each combination of fields turns into an independent matrix element. Since these are standard procedures (see e.g. \cite{Boer:2003cm, Balitsky:2020jzt, Vladimirov:2021hdn}) we do not present them in details, and we write directly the final expression. 

The bare factorization theorem reads
\begin{eqnarray}\label{qTMD:bare-position}
\widetilde{\Omega}_{\text{bare}}^{[\Gamma]}(y)&=&
\Psi (b)\widehat{C}^\dagger_1\widehat{C}_1 \Big[
\widetilde{\Phi}_{11}^{\llbracket\Gamma\rrbracket}(\ell,b)-\frac{1}{2}\frac{\partial_\mu}{\partial_+}\widetilde{\Phi}_{11}^{\llbracket\gamma^\mu\gamma^+\Gamma+\Gamma \gamma^+\gamma^\mu\rrbracket}(\ell,b)\Big]
\\\nn &&
+\frac{i}{2}\Psi(b)\int_{s\infty}^0 d\sigma \frac{1}{\partial_+}\Big[
\widehat{C}_2^\dagger \widehat{C}_1\widetilde{\Phi}_{\mu,21}^{\llbracket\gamma^\mu\gamma^+\Gamma\rrbracket}(\ell,\ell+\sigma,0,b)
+
\widehat{C}_1^\dagger \widehat{C}_2\widetilde{\Phi}_{\mu,12}^{\llbracket\Gamma\gamma^+\gamma^\mu\rrbracket}(\ell,\sigma,0,b)\Big]
\\\nn &&
+\frac{i}{2}\int_{-\infty}^0 d\sigma  \frac{1}{\partial_+}\Big[
\Psi_{\mu,21}(\sigma,b)\widehat{C}_{2v}^\dagger \widehat{C}_{1}\widetilde{\Phi}_{11}^{\llbracket\gamma^\mu\gamma^+\Gamma\rrbracket}(\ell,b)
+
\Psi_{\mu,12}(\sigma,b)
\widehat{C}_{1}^\dagger \widehat{C}_{2v}\widetilde{\Phi}_{11}^{\llbracket\Gamma\gamma^+\gamma^\mu\rrbracket}(\ell,b)\Big],
\end{eqnarray}
where indices $\mu$ are transverse, and $\partial_\mu=\partial/\partial b^\mu$, $\partial_+=\partial/\partial \ell$. Here, all distributions are  bare distributions. To write the factorization theorem in this form, we used the total-shift invariance of forward matrix elements.

The functions $\Psi$ and $\widetilde{\Phi}$ parametrize the nonperturbative parts of the factorization formula. Namely, the functions $\widetilde{\Phi}$ are TMD distributions in position space
\begin{eqnarray}\label{def:TMD11}
\widetilde{\Phi}^{[\Gamma]}_{11}(\ell,b)&=&\langle P,S|T\{\bar q(\ell n+b)\frac{\Gamma}{2}q(0)\}|P,S\rangle,
\\\label{def:TMD21}
\widetilde{\Phi}_{\mu,21}^{[\Gamma]}(z_1,z_2,z_3,b)&=&g\langle P,S|T\{\bar q(z_1 n+b) F_{\mu+}(z_2n+b)\frac{\Gamma}{2}q(z_3n)\}|P,S\rangle,
\\\label{def:TMD12}
\widetilde{\Phi}_{\mu,12}^{[\Gamma]}(z_1,z_2,z_3,b)&=&g\langle P,S|T\{\bar q(z_1 n+b) \frac{\Gamma}{2}F_{\mu+}(z_2n)q(z_3n)\}|P,S\rangle,
\end{eqnarray}
where we omit the Wilson lines along direction $n$ that connect all fields and continue to $s\infty n$. The double bracket on the Dirac structure impose the projection to the good component only, i.e.
\begin{eqnarray}
\widetilde{\Phi}^{\llbracket\Gamma\rrbracket}=\widetilde{\Phi}^{[\frac{\gamma^+\gamma^-}{2}\Gamma\frac{\gamma^-\gamma^+}{2}]}=\frac{1}{2}\Tr\(\widetilde{\Phi}\frac{\gamma^+\gamma^-}{2}\Gamma\frac{\gamma^-\gamma^+}{2}\).
\end{eqnarray}
The functions $\Psi$ are defined as 
\begin{eqnarray}
\Psi(b)&=&\langle 0|\frac{\Tr}{N_c}[-\bar n\infty+b,b]H(b) H^\dagger(0)[0,-\bar n\infty]|0\rangle,
\\
\Psi_{\mu,12}(z,b)&=&\langle 0|\frac{\Tr}{N_c}[-\bar n\infty+b,b]H(b) H^\dagger(0)[0,z\bar{n}]F_{\mu-}[z\bar{n},-\bar n\infty]|0\rangle,
\\
\Psi_{\mu,21}(z,b)&=&\langle 0|\frac{\Tr}{N_c}[-\bar n\infty+b,z\bar{n}+b]F_{\mu-}[z\bar{n}+b,b]H(b) H^\dagger(0)[0,-\bar n\infty]|0\rangle.
\end{eqnarray}
and are vacuum expectation values of close Wilson loops. All definitions implicitly contain transverse links that connect the far ends of the Wilson lines. These links are required to make the definition strictly gauge invariant. 

Let us stress that the TMD-distributions (\ref{def:TMD11}, \ref{def:TMD21}, \ref{def:TMD12}) are independent nonperturbative functions. Each of them obeys a separate closed evolution equation. The indexing 11, 12, and 21 refers to the TMD-twist of this operators \cite{Vladimirov:2021hdn}. So, the operator of TMD distribution $\Phi_{11}$ consists of product of two (so-called semi-compact) operators of twist-one (good component of quark field with an attached light-like Wilson line), and it has TMD-twist-(1,1). The operators for TMD distributions $\Phi_{12}$ and $\Phi_{21}$ consist of operators twist-one and twist-two (good component of quark field and good component of the gluon field with attached light-like Wilson line). Therefore, they have TMD-twist-(1,2) and TMD-twist-(2,1), respectively. In the following, we refer to TMD distributions of TMD-twist-(1,2) and TMD-twist-(2,1), simply as TMD distributions of twist-three.

The $\Psi$-functions do not have definite twist, because the twist of the Wilson line $H_v$ is not defined. Nonetheless, their enumeration follows the same pattern for analogy. These function are nonperturbative objects with autonomous evolution.

\subsection{Bare qTMD correlator at NLP (momentum-fraction space)}
\label{sec:bare-momentum}

Taking the Fourier transformation with respect to $\ell$, we obtain the momentum-fraction representation (\ref{def:Fourier-OMEGA}) for the factorization theorem, namely
\begin{eqnarray}\label{qTMD:bare-momentum}
\Omega^{[\Gamma]}_{\text{bare}}(x,b)&=&
\Psi (b)C_1(-x)C_1(x) 
\Phi_{11}^{\llbracket\Gamma\rrbracket}(x,b)
\\\nn && 
+\frac{i \partial_\mu}{2x P_+} \Psi (b)C_1(-x)C_1(x) \Phi_{11}^{\llbracket\gamma^\mu\gamma^+\Gamma+\Gamma \gamma^+\gamma^\mu\rrbracket}(x,b)
\\\nn &&
+\frac{i}{2xP_+}\Psi(b)\int \frac{[dx]}{x_2-is0}\Big[
\delta(x-x_3)C_2(x_2,x_1) C_1(x_3)\Phi_{\mu,21}^{\llbracket\gamma^\mu\gamma^+\Gamma\rrbracket}(x_{1,2,3},b)
\\&&\nn
\qquad\qquad
+
\delta(x+x_1)C_1(x_1) C_2(x_2,x_3) \Phi_{\mu,12}^{\llbracket\Gamma\gamma^+\gamma^\mu\rrbracket}(x_{1,2,3},b)\Big]
\\\nn &&
+\frac{1}{2xP_+}\int_{-\infty}^0 d\sigma  \Big[
\Psi_{\mu,21}(\sigma,b)C_{2v}(-x) C_{1}(x)\Phi_{11}^{\llbracket\gamma^\mu\gamma^+\Gamma\rrbracket}(x,b)
\\&&\nn
\qquad\qquad
+
\Psi_{\mu,12}(\sigma,b)
C_{1}(-x) C_{2v}(x)\Phi_{11}^{\llbracket\Gamma\gamma^+\gamma^\mu\rrbracket}(x,b)\Big].
\end{eqnarray}
where $(x_{1,2,3},b)$ is the short notation for $(x_1,x_2,x_3,b)$. The integral measure is defined as
\begin{eqnarray}\label{def:[dx]}
\int [dx]=\int_{-1}^1 dx_1dx_2dx_3\delta(x_1+x_2+x_3),
\end{eqnarray}
which is the consequence of  momentum conservation. 

The TMD distributions in the momentum-fraction space are defined as follows
\begin{eqnarray}
\widetilde{\Phi}_{11}^{[\Gamma]}(\ell,b)&=&P^+\int_{-1}^1 dx e^{ix \ell P^+}\Phi_{11}^{[\Gamma]}(x,b),
\\
\widetilde{\Phi}^{[\Gamma]}_{\mu,ij}(z_1,z_2,z_3,b)&=&(P^+)^2\int[dx] e^{-i (z_1x_1+z_2x_2+z_3x_3) P^+}\Phi^{[\Gamma]}_{\mu, ij}(x_1,x_2,x_3,b),
\end{eqnarray}
where $ij$ is 12 or 21. We stress the ``minus'' sign in the definition of momentum-fractions for twist-three distributions. Such definition provides a ``natural'' partonic interpretation \cite{Jaffe:1983hp}.

The coefficient functions obtained are
\begin{eqnarray}\label{C1:bare-mom}
C_1(x)&=&1+2a_sC_F\frac{1-\epsilon}{1-2\epsilon}\Gamma(-\epsilon)\Gamma(2\epsilon)\(\frac{-v^2}{4}\)^{\epsilon}\frac{1}{(isx (vP))^{2\epsilon}}+\mathcal{O}(a_s^2),
\\\label{C2:bare-mom}
C_2(x_2,x_3)&=&1+2a_s\Gamma(-\epsilon)\Gamma(2\epsilon)\(\frac{-v^2}{4}\)^{\epsilon}\frac{1}{(is(x_2+x_3) (vP))^{2\epsilon}}\Bigg\{
C_F\frac{1-3\epsilon}{1-2\epsilon}
\\\nn &&
-\(C_F-\frac{C_A}{2}\)\frac{1}{1-2\epsilon}\frac{x_2+x_3}{x_2}\(1-\(\frac{x_2+x_3-is0}{x_3-is0}\)^{2\epsilon}\)
\\\nn &&
+C_A\frac{\epsilon}{1-2\epsilon}\frac{x_2+x_3}{x_3}\(1-\(\frac{x_2+x_3-is0}{x_2-is0}\)^{2\epsilon}\)\Bigg\}+\mathcal{O}(a_s^2)
,
\end{eqnarray}
The first argument of $C_2$ is related to the momentum of gluon, and the second is the momentum of quark or anti-quark. 

We emphasize that the signs of momentum-fractions are not restricted. The TMD distributions and qTMD correlators are defined for positive and negative values of the momentum fractions. TMD distributions of twist-two with the negative values of $x$ are associated with the anti-parton distributions. TMD distributions of twist-three have a more involved interpretation. Three momentum fractions $x_1$, $x_2$ and $x_3$ are related to each other by momentum conservation $x_1+x_2+x_3=0$, which is reflected in the delta-function in the integral measure (\ref{def:[dx]}). There are six combinations of signs for $x$'s. Each combination has a separate partonic interpretation for $\Phi_{\mu,12}$ and $\Phi_{\mu,21}$ \cite{Rodini:2022wki}. The important point is that different ranges of $x$'s are mixed in the integral convolutions with coefficient functions or with evolution kernel. In the formulas above, the restrictions for the integration domains should be found for each particular term resolving delta-functions. For example, the integral that appears in the second line of (\ref{qTMD:bare-momentum}) explicitly reads (for $x>0$)
\begin{eqnarray}\label{example-integral}
&&\int \frac{[dx]}{x_2-is0}\delta(x-x_3)C_2 (x_2,x_1)C_1(x_3)\Phi_{\mu,21}^{\llbracket \gamma^\mu\gamma^+\Gamma\rrbracket}(x_1,x_2,x_3,b)
\\\nn&&=
\int_{-1}^{1-x} \frac{dx_2}{x_2-is0}C_2 (x_2,-x-x_2)C_1(x)\Phi_{\mu,21}^{\llbracket \gamma^\mu\gamma^+\Gamma\rrbracket}(-x-x_2,x_2,x,b).
\end{eqnarray}
Here, the integration involves both positive and negative values of $x_2$. At the point $x_2=0$ the integrand is singular. Therefore, the $is0$ prescriptions are of utter importance. They are responsible for a number of effects discussed in the following section.

\section{Factorization theorem in the physical terms}
\label{sec:bare->physical}

The derivation of the bare form of the factorization theorem is only the halfway point in the derivation of the final expression. To obtain a presentation suitable for practical applications, one needs to perform several manipulations and combine together different elements. Some of these iterations require additional computations. In this section, we collect the key points of this procedure and describe the process of deriving the factorization expression in the physical terms. The well-known elements (such as the recombination of rapidity divergences) are discussed very briefly, whereas the novel aspects are presented in some detail.

Structurally, the TMD factorization at NLP is more involved than at LP. It contains a larger number of details to be treated. The first point to address is the cancellation of divergences in-between different elements of the formula. The NLP TMD factorization contains the following combinations of divergences
\begin{itemize}
\item The infrared (IR) divergences of the coefficient function are presented as $1/\epsilon$-terms in the bare expressions (\ref{C1:bare-mom}, \ref{C2:bare-mom}). These poles are canceled by the ultraviolet (UV) renormalization constants of TMD distributions and $\Psi$-functions. The cancellation of $1/\epsilon$-terms is, however, not complete. The leftover is the UV pole corresponding to the renormalization of currents $J_v$. Since the TMD distributions of distinct TMD-twists are independent nonperturbative functions, such cancellation must happen individually for each term of the bare expression (\ref{qTMD:bare-position}). Schematically, one should observe that each term of factorized expression satisfies
\begin{eqnarray}
Z_{J}^{-2} C_i^\dagger C_j Z_{ij}^{\text{TMD}} Z_{\Psi}=\text{finite},
\end{eqnarray}
where $Z_J$ is the renormalization constant for $J_v$, $Z_{ij}^{\text{TMD}}$ is the renormalization constant for $\Phi_{ij}$, and $Z_{\Psi}$ is the renormalization constant for the corresponding $\Psi$-function. 
\item The rapidity divergences of TMD operator and $\Psi$-functions are canceled by the soft factor $S(y)$ (\ref{def:func0}). At LP and NLP the soft factor is ordinary TMD soft factor \cite{Ebert:2021jhy}. The cancellation of rapidity divergences for NLP operators has been demonstrated explicitly at NLO in ref.\cite{Vladimirov:2021hdn}. In the present context, the only difference from computation in ref. \cite{Vladimirov:2021hdn} are the $\Psi$-functions. The function $\Psi$ has been studied in refs.\cite{Ebert:2019okf, Vladimirov:2020ofp}, and its rapidity divergence is identical to $\Phi_{11}$. The rapidity divergences of $\Psi_{\mu,12}$ and $\Psi_{\mu,21}$ functions are the same as for $\Psi$ at NLO, which can be checked by direct computation. Therefore, all rapidity divergent factors cancel at NLP just as they cancel at LP \cite{Ebert:2019okf, Ji:2019ewn, Vladimirov:2020ofp}. We do not provide a deeper discussion.
\item The integrals over $\sigma$ for NLP terms (\ref{qTMD:bare-position}) are divergent at $\sigma \to s\infty$. In momentum-fraction space (\ref{qTMD:bare-momentum}), this divergence transforms to the divergence at $x_2\to0$. These divergences are called ``special'' rapidity divergences \cite{Rodini:2022wki}. Special rapidity divergences are implicit and cancel between collinear and $v$-collinear sectors (second and third lines in eqn. (\ref{qTMD:bare-position})). To make the factorization formula finite term-by-term, one defines physical TMD distributions by adding (and subtracting in the factorized formula) specific divergent pieces. Special rapidity divergences are specific for power corrections, i.e., they are trivially absent at LP.
\end{itemize}
The computation of rapidity divergences is performed in the $\delta$-regularization defined in refs.\cite{Echevarria:2015byo, Echevarria:2016scs}. 

As a result of these procedures, one obtains the finite expression for the qTMD correlator with each nonperturbative element satisfying an evolution equation (given in sec.\ref{sec:intemidiate-form}). Even so, the expression is still not very practical. It contains a combination of terms with different imaginary parts. The last step is to resolve the complex structure and present the formula in a directly usable form.

In the following sections, we discuss in particular detail these procedures. The final expression for the NLP factorization of the qTMD correlator is given in sec.\ref{sec:fac-final}.

\subsection{TMD distributions: properties and evolution}
\label{sec:TMD-dist}

The bare TMD distributions are defined in eqns.(\ref{def:TMD11}, \ref{def:TMD21}, \ref{def:TMD12}). Their renormalization and evolution properties are known. For the detailed description of twist-three TMD distribution we refer to ref.\cite{Rodini:2022wki}. In this section, we briefly summarize the features that are important for the present work.

All TMD distributions are renormalized by three factors. Two UV renormalization constant (one for each semi-compact operator), and the rapidity renormalization factor. For the present case, we have
\begin{eqnarray}\nn
\Phi^{[\Gamma]}_{11,\text{bare}}(x,b)&=& R(b^2)Z_{U1}(-x)Z_{U1}(x)\Phi^{[\Gamma]}_{11}(x,b;\mu,\zeta),
\\\label{TMD:renorm}
\Phi^{[\Gamma]}_{\mu,21,\text{bare}}(x_1,x_2,x_3,b)&=& R(b^2)Z_{U2}(x_2,x_1)Z_{U1}(x_3)\otimes\Phi^{[\Gamma]}_{\mu,12}(x_1,x_2,x_3,b;\mu,\zeta),
\\\nn
\Phi^{[\Gamma]}_{\mu,12,\text{bare}}(x_1,x_2,x_3,b)&=& R(b^2)Z_{U1}(x_1)Z_{U2}(x_2,x_3)\otimes\Phi^{[\Gamma]}_{\mu,12}(x_1,x_2,x_3,b;\mu,\zeta),
\end{eqnarray}
where $Z_{U1}$ and $Z_{U2}$ are the UV renormalization constants for the twist-1 and twist-2 semi-compact constituents of the TMD operator \cite{Rodini:2022wki}. The factor $R$ is the renormalization constant for the rapidity divergence \cite{Chiu:2012ir, Vladimirov:2017ksc}. Loosely speaking, $R=1/\sqrt{S}$ \cite{Echevarria:2012js, Collins:2011zzd, Vladimirov:2017ksc}. The scales $\mu$ and $\zeta$ are the scales of UV and rapidity renormalization, respectively. The symbol $\otimes$ denotes the integral convolution in $x$'s between $Z_{U2}$ and the TMD distribution. 

The rapidity divergences are associated with the light-cone directions, and thus their renormalization introduces the non-boost-invariant scales $\nu^\pm$. The UV renormalization constant also have dependence on light-cone direction, which appear in the collinear-divergent part and scales with the common momentum $q^\pm$ passing through the operator. The soft factor $S(y)$ (\ref{def:func0}) cancels the rapidity and collinear divergences. It also restores the boost-invariance through the introduction of the boost-invariant scales for rapidity evolution:
\begin{eqnarray}
\label{Zeta_ZetaBar_def}
\zeta=2(q^+)^2\frac{\nu^-}{\nu^+},\qquad 
\bar \zeta=2(v^-\mu)^2\frac{\nu^+}{\nu^-}.
\end{eqnarray}
where the $\bar\zeta$ can only be proportional to $\mu^2$, since no other hard scale is present in the $v$-collinear sector. The resulting renormalization factors are called subtracted, and depend on $\zeta$. Details on the whole procedure can be found in refs.\cite{Echevarria:2012js, Chiu:2012ir, Collins:2011zzd, Vladimirov:2017ksc}.

The NLO expression for the subtracted renormalization constant $Z^{\text{sub.}}_{U1}$ is well-known \cite{Aybat:2011zv, Echevarria:2011epo}:
\begin{eqnarray}\label{def:ZU1}
Z^{\text{sub.}}_{U1}&=&1+\frac{a_s}{\epsilon}C_F\(\frac{1}{\epsilon}+\frac{3}{2}+\ln\(\frac{\mu^2}{\zeta}\)-2\log\(i s s_x\)\)+\mathcal{O}(a_s^2),
\end{eqnarray}
where  $s_x=\sign(x)$, and we included the, usually neglected, imaginary part. This imaginary part is inessential for standard Drell-Yan/SIDIS NLP factorization (see ref.\cite{Vladimirov:2021hdn}) where $s_x$ and $s$ are fixed by the process kinematics. In the context of qTMD factorization, the signs $s_x$ and $s$ are not fixed, and thus $\log(iss_x)$ is important and one should keep track of these terms explicitly. The expression for $Z_{U2}^{\text{sub.}}$ is complicated \cite{Vladimirov:2021hdn, Rodini:2022wki}, and it is not important for the present computation. In the present context, the $Z^{\text{sub.}}_{U2}$ enters the integral (\ref{qTMD:bare-momentum}), and thus the convolution structure can be simplified. One has
\begin{eqnarray}
\int \frac{[dx]}{x_2-is0} Z^{\text{sub.}}_{U2}(x_2,x_3)\otimes U(x_2,x_3)=
\int \frac{[dx]}{x_2-is0} Z^{(0){\text{sub.}}}_{U2}(x_2,x_3) U(x_2,x_3),
\end{eqnarray}
where $U$ is a test function and 
\begin{eqnarray}\label{def:ZU2}
&&Z^{(0){\text{sub.}}}_{U2}(x_2,x_3)=1
\\\nn && +\frac{a_s}{\epsilon}\Big[
C_F\(\frac{1}{\epsilon}-\frac{1}{2}+\ln\(\frac{\mu^2}{\zeta}\)-2\log\(i s s_x\)\)+2\(C_F-\frac{C_A}{2}\)\frac{x_2+x_3}{x_2}\ln\(\frac{x_2+x_3-is0}{x_3-is0}\)\Big]
\\\nn &&
+\mathcal{O}(a_s^2).
\end{eqnarray}
The derivation of $Z^{(0)\text{sub}}_{U2}$ can be found in ref.\cite{Vladimirov:2021hdn}. 

The renormalized TMD distributions satisfy a pair of evolution equations (\ref{evol:UV}, \ref{evol:zeta}). The scaling with respect to $\mu$ reads
\begin{eqnarray}\label{evol:UV}
\mu^2\frac{d}{d\mu^2}\Phi^{[\Gamma]}_{ij}(\{x\}_i,\{x\}_j,b;\mu,\zeta)=\(\gamma^\dagger_i(\{x\}_i,\mu,\zeta)+\gamma_j(\{x\}_j,\mu,\zeta)\)\otimes\Phi^{[\Gamma]}_{ij}(\{x\}_i,\{x\}_j,b;\mu,\zeta),
\end{eqnarray}
where $\{x\}_n$ indicates a collection of $n$ momentum fractions, $\gamma_i$ are anomalous dimensions, and $\otimes$ is the integral convolution in $x$'s.  In the present work we need only anomalous dimensions $\gamma_1$ and $\gamma_2$. Both anomalous dimensions have complex parts. For anomalous dimension $\gamma_1$, the complex phase accumulates the full dependence on the momentum fraction. In the case of TMD distributions of twist-two these complex parts cancel entirely in the sum and the result is the well-known expression, which at LO is
\begin{eqnarray}
\gamma^\dagger_1(x,\mu,\zeta)+\gamma_1(x,\mu,\zeta)&=&2a_s C_F\(\ln\(\frac{\mu^2}{\zeta}\)+3\)+\mathcal{O}(a_s^2).
\end{eqnarray}
The anomalous dimension $\gamma_2$ is cumbersome, which leads to an involved expression for the evolution equation already at LO \cite{Rodini:2022wki}. 

The evolution with respect to scale $\zeta$ reads
\begin{eqnarray}\label{evol:zeta}
\zeta \frac{d}{d\zeta}\Phi^{[\Gamma]}_{ij}(\{x\}_i,\{x\}_j,b;\mu,\zeta)
=
-\mathcal{D}(b,\mu)\Phi^{[\Gamma]}_{ij}(\{x\}_i,\{x\}_j,b;\mu,\zeta),
\end{eqnarray}
where $\mathcal{D}$ is the Collins-Soper kernel \cite{Collins:1981uk}. The equation (\ref{evol:zeta}) is valid for $ij$=11, 12, 21. The Collins-Soper kernel is a nonperturbative function. UV and rapidity anomalous dimensions satisfy the integrability condition \cite{Chiu:2012ir, Scimemi:2018xaf}
\begin{eqnarray}\label{integrability}
-\zeta\frac{d}{d\zeta}\(\gamma_i(\{x\}_i,\mu,\zeta)+\gamma_j(\{x\}_j,\mu,\zeta)\)=\mu^2 \frac{d}{d\mu^2}\mathcal{D}(b,\mu)=\frac{\Gamma_{\text{cusp}}(\mu)}{2},
\end{eqnarray}
where $\Gamma_{\text{cusp}}$ is the cusp anomalous dimension.

At small values of $b$ the TMD distributions can be computed in the terms of collinear PDFs by means of operator product expansion (OPE). Herewith, there is no relation between the TMD-twist of TMD distribution and the collinear twist of PDF. So, for TMD-twist-two distributions, the leading term of OPE has the form $\Phi_{11}(x,b)\sim C(x,\ln(b))\otimes f(x)$ where $C$ is a perturbative coefficient, and $f(x)$ is a collinear distribution of twist-two or -three, see examples in refs.\cite{Echevarria:2016scs,Scimemi:2019gge}, and complete analysis in ref.\cite{Moos:2020wvd}. For the TMD distributions of TMD-twist-three the situation is more involved, since the leading term of OPE can be singular. General structure of OPE has the form
\begin{eqnarray}\label{small-b-tw-3}
\lim_{b\to 0}\Phi_{\mu,12}^{[\Gamma]}(x,b)\sim \frac{a_s b_\mu}{b^2}C^{[\Gamma]}_{1}(x,\ln(b))\otimes f_1(x)+C^{[\Gamma]}_{\mu,2}(x,\ln(b))\otimes f_2(x)+\mathcal{O}(b),
\end{eqnarray}
and similar for $\Phi_{21}$. Here, $C_1$ and $C_2$ are perturbative coefficient functions, $\otimes$ is an integral convolution, and $f_{1}$ are collinear distributions of collinear twist-two, and $f_2$ are collinear distributions and twist-three and higher. Note, that the $b^{-1}$-term is $a_s$-suppressed. The explicit expressions for coefficient functions $C_1$ can be found in appendix C of ref.\cite{Rodini:2022wki}. In this way, the factorization formula (\ref{qTMD:bare-momentum}) has the $\mathcal{O}(a_s/b(vP))$ behaviour in the perturbative approximation. Partially, it comes from $\partial_\mu \Phi_{11}$ (due to the derivative of $\ln(b)$ at NLO), and partially, from twist-three terms (\ref{small-b-tw-3}). It also shows that in order to receive the NLP TMD factorization from the resummation approach, one must take into account collinear twist-two and twist-three operators (at least).

Finally, we have to address the emergence of the special rapidity divergences. As it is discussed in sec.\ref{sec:bare-momentum}, the point $x_2=0$ is the singular point of the factorized expression. If the TMD distribution is continuous at $x_2=0$, the integral around this point will only produce an imaginary part. However, twist-three distributions are generally discontinuous at $x_2=0$. Therefore, the integrals of type (\ref{example-integral}) are divergent.

The divergences of integrals of type (\ref{example-integral}) are rapidity divergences. They are a different type of rapidity divergences compared to the ordinary one of the TMD operator that are renormalized by the factor $R$ (\ref{TMD:renorm}), and for that reason are called ``special rapidity divergences''. Special rapidity divergences can be computed explicitly order-by-order in perturbation theory. In ref.\cite{Rodini:2022wki}  it is shown that the LO special rapidity divergence for TMD distributions of twist-three is
\begin{eqnarray}\label{sp.rap.div1}
&&\int \frac{[dx]}{x_2-is \frac{\delta^+}{q^+}}\delta(x-x_3)\Phi^{[\Gamma]}_{\mu,21}(x_{1,2,3},b)=-
\ln\(\frac{\delta^+}{q^+}\)\partial_\mu\mathcal{D}(b)
\Phi^{[\Gamma]}_{11}(x,b)+\text{fin.terms}+\mathcal{O}(a_s^2),
\\\nn
&&\int \frac{[dx]}{x_2-is \frac{\delta^+}{q^+}}\delta(x+x_1)\Phi^{[\Gamma]}_{\mu,12}(x_{1,2,3},b)=-
\ln\(\frac{\delta^+}{q^+}\)\partial_\mu\mathcal{D}(b)
\Phi^{[\Gamma]}_{11}(x,b)+\text{fin.terms}+\mathcal{O}(a_s^2),
\end{eqnarray}
where $\delta^+$ is the $\delta$-regulator,  $q^+$ is the momentum passing through the Wilson line, and $\mathcal{D}$ is the Collins-Soper kernel. 

A feature of the special rapidity divergence is that it is proportional to the TMD distributions of twist-two. This can be used to re-define TMD distributions in a controllable and systematic manner. One defines \textit{physical} TMD distributions by subtracting a precomputed finite term such that the integrals of type (\ref{sp.rap.div1}) are finite. The factorized expression spelled in the terms of physical TMD distributions is term-by-term finite. We define
\begin{eqnarray}\label{TMD:phys}
\mathbf{\Phi}^{[\Gamma]}_{\mu,ij}(x_{1,2,3},b;\mu,\zeta)
&=&
\Phi^{[\Gamma]}_{\mu,ij}(x_{1,2,3},b;\mu,\zeta)
-[\mathcal{R}_{ij}\otimes \Phi_{11}]_\mu^{[\Gamma]}(x_{1,2,3},b;\mu,\zeta),
\end{eqnarray}
where $ij$ is 12 or 21, and $[\mathcal{R}\otimes \Phi]$ is a convolution of $\Phi_{11}$ and perturbative function. The explicit form of $[\mathcal{R}\otimes \Phi]$ at LO can be found in ref.\cite{Rodini:2022wki}. We stress that the definition (\ref{TMD:phys}) is made on the renormalized TMD distributions and it does not change the evolution equations for them.

\subsection{$\Psi$-functions: properties and evolution}
\label{sec:Psi-distr}

The factorization theorem for qTMD correlator contains new objects -- $\Psi$-functions. To our best knowledge these functions are specific to the factorization of the qTMD correlator. On the one hand they are similar to the ordinary TMD distributions, but with the parton field replaced by Wilson lines along direction $v$ (or equivalently by the field $H$ (\ref{def:H})). For that reason, some of the properties of $\Psi$-functions are analogous to the properties of TMD distribution (for example, double-scale evolution). On the other hand, they are similar to correlators of two heavy-quark fields, and some of their properties could be deduced by analytical continuation to $v^2<0$.  We summarize important properties of $\Psi$-functions in this section.

The renormalization of $\Psi$ functions is
\begin{eqnarray}\nn
\Psi_{\text{bare}}(b)&=& R(b^2)Z_{W}(b)Z_{\Psi1}^2\Psi(b;\mu,\zeta),
\\\label{Psi:renorm}
\Psi_{\mu,21,\text{bare}}(\sigma,b)&=& R(b^2)Z_{W}(b)Z_{\Psi2}Z_{\Psi1}\otimes \Psi_{\mu,21}(\sigma,b;\mu,\zeta),
\\\nn
\Psi_{\mu,12,\text{bare}}(\sigma,b)&=& R(b^2)Z_{W}(b)Z_{\Psi1}Z_{\Psi2}\otimes \Psi_{\mu,12}(\sigma,b;\mu,\zeta),
\end{eqnarray}
where factor $R$ is the rapidity renormalization factor same as in the TMD distribution case (\ref{TMD:renorm}). The factors $Z_{\Psi1}$ and $Z_{\Psi2}$ are the UV renormalization (in the light-cone gauge) of operators $H$ and $HF_{\mu-}$  correspondingly. 

The factor $Z_{W}$ represents the renormalization of the staple finite-size contour in the direction $v$. This factor is the same for $\Psi$-function and for qTMD correlator (\ref{def:quasiTMD}), because this part   passes intact from the initial definition to the factorized form. In our computation of the coefficient function we did not include the computation of self-energies for $v$-directed Wilson lines, which are totally absorbed into the factor $Z_W$. All power-unsuppressed differences between the finite (but large) and infinite $L$ are accumulated in this factor. For a more detailed discussion on the order of limits in the factorization for qTMD correlator see ref.\cite{Ebert:2022fmh}.

The $Z_{\Psi1}$ and $Z_{\Psi2}$ contains the collinear divergences in the same way as factors $Z_{U1}$ and $Z_{U2}$. They are removed, along with  the factor $R$,  by the soft factor in the same way as for TMD distributions. The resulting subtracted renormalization constants depend only boost-invariant variables (\ref{evol:zeta}). 
The factor $Z_{\Psi1}$ has been computed\footnote{
In ref.\cite{Vladimirov:2020ofp} the final expression for $Z_{\Psi1}$ contains a mistake, due to the different definition of renormalization factor $Z_J$ (\ref{def:ZJ}) that is taken from ref.\cite{Chetyrkin:2003vi}. Here, the mistake is corrected.
} in ref.\cite{Vladimirov:2020ofp}, and it reads
\begin{eqnarray}\label{def:Zpsi}
Z^{\text{sub.}}_{\Psi1}&=&1+\frac{a_s}{\epsilon}C_F\(1+\ln\(\frac{\mu^2}{\zeta(\mu^2)}\)\)+\mathcal{O}(a_s^2),
\end{eqnarray}
where we stress that the rapidity scale $\zeta$ is actually a function of the UV scale $\mu$ by the way it is introduced in eq.\eqref{Zeta_ZetaBar_def}. The dependence is such that
\begin{equation}
\mu^2 \frac{\partial}{\partial\mu^2} \ln\( \frac{\mu^2}{\zeta(\mu^2)} \) = 0.
\end{equation}
Keeping this dependence explicit is important to have $\gamma_\Psi=d\ln Z_{\Psi1}^{\text{sub.}}/d\ln \mu^2$ finite, since $Z^{\text{sub.}}_{\Psi1}$ does not contain double-pole in $\epsilon$. The expression for $Z^{\text{sub.}}_{\Psi2}$ contains a convolution in the position of gluon field, and is not important for the present case since it does not appear in the factorized expression (\ref{qTMD:bare-position}). The combinations that appear are the ``zeroth'' moments of the functions $\Psi_{\mu,12}$ and $\Psi_{\mu,21}$. We introduce the special notation for them
\begin{eqnarray}\label{def:Psi0}
\Psi^{(0)}_{\mu,21}(b)=\int_{-\infty}^0 d\sigma \Psi_{\mu,21}(\sigma,b),
\qquad
\Psi^{(0)}_{\mu,12}(b)=\int_{-\infty}^0 d\sigma \Psi_{\mu,12}(\sigma,b).
\end{eqnarray}
The renormalization of $\Psi^{(0)}$ is the same as for $\Psi$, but with $Z_{\Psi2}$ replaced by $Z_{\Psi2}^{(0)}$. The factor $Z_{\Psi2}^{(0)}$ is multiplicative. We found that at NLO
\begin{eqnarray}
Z_{\Psi2}^{(0)\text{sub.}}=Z_{\Psi1}^{\text{sub.}}+\mathcal{O}(a_s^2).
\end{eqnarray}
This relation could be a consequence of soft-gluon theorems, similarly to the relation between coefficient functions $C_{2v}$ and $C_1$. However, for the moment, we cannot state it exactly.

In the complete analogy to the $x_2$-integral with TMD distributions, the integrals (\ref{def:Psi0}) exhibit the special rapidity divergence at $\sigma\to-\infty$. The one-loop computation yields
\begin{eqnarray}\label{rap.div:Psi}
&&\int_{-\infty}^{0}d\sigma e^{\delta^- \sigma}\Psi_{\mu,21}(\ell,b)=
i\ln\(\frac{\delta^-}{q^-}\)\partial_\mu\mathcal{D}(b)
\Psi(b)+\text{fin.terms}+\mathcal{O}(a_s^2),
\\\nn
&&\int_{-\infty}^{0}d\sigma e^{\delta^- \sigma}\Psi_{\mu,12}(\ell,b)=
i\ln\(\frac{\delta^-}{q^-}\)\partial_\mu\mathcal{D}(b)
\Psi(b)+\text{fin.terms}+\mathcal{O}(a_s^2).
\end{eqnarray}
Using this expression, we define the finite functions $\mathbf{\Psi}^{(0)}$ as
\begin{eqnarray}\label{Psi:phys}
\mathbf{\Psi}^{(0)}_{\mu,ij}(b;\mu,\zeta)=\Psi^{(0)}_{\mu,ij}(b;\mu,\zeta)-i\ln\(\frac{\delta^-}{\nu^-}\)\partial_\mu\mathcal{D}(b,\mu)
\Psi(b;\mu,\zeta),
\end{eqnarray}
with $ij$ being $12$ or $21$.

The evolution equations for the $\Psi$-functions are
\begin{eqnarray}
\mu^2 \frac{d}{d\mu^2}\Psi(b;\mu,\zeta)&=&
2\gamma_{\Psi}\Psi(b;\mu,\zeta),
\\
\mu^2 \frac{d}{d\mu^2}\mathbf{\Psi}^{(0)}_{\mu,ij}(b;\mu,\zeta)&=&
(\gamma_{\Psi2}+\gamma_{\Psi})\mathbf{\Psi}^{(0)}_{\mu,ij}(b;\mu,\zeta),
\end{eqnarray}
where $ij$ is $12$ or $21$, and
\begin{eqnarray}\label{def:AD:PSI}
\gamma_{\Psi}&=&a_sC_F\(\ln\(\frac{\mu^2}{\zeta(\mu^2)}\)+1\)+\mathcal{O}(a_s^2),\qquad
\gamma_{\Psi2}=\gamma_{\Psi}+\mathcal{O}(a_s^2).
\end{eqnarray}
Note that there should also be a part of anomalous dimension associated with the $Z_W$ constant. Here we ignore it, assuming that the renormalization of contour is made on a separate scale. The evolution with respect to the rapidity scale is the same as for TMD distributions
\begin{eqnarray}\label{Psi:rap.evol.}
\zeta \frac{d}{d\zeta}\Psi(b;\mu,\zeta)&=&-\mathcal{D}(b,\mu)\Psi(b;\mu,\zeta),
\qquad
\zeta \frac{d}{d\zeta}\mathbf{\Psi}^{(0)}_{\mu,ij}(b;\mu,\zeta)=
-\mathcal{D}(b,\mu)\mathbf{\Psi}^{(0)}_{\mu,ij}(b;\mu,\zeta).
\end{eqnarray}
The UV anomalous dimensions $\gamma_{\Psi}$ also satisfy the integrability condition (\ref{integrability}).

The functions $\Psi^{(0)}_{\mu,12}$ and $\Psi^{(0)}_{\mu,21}$ are not independent. Using discrete symmetries one finds
\begin{eqnarray}
\mathbf{\Psi}^{(0)}_{\mu,12}(b)=(\mathbf{\Psi}^{(0)}_{\mu,21}(-b))^*=\mathbf{\Psi}^{(0)}_{\mu,21}(b).
\end{eqnarray}

At small-$b$ the $\Psi$-functions are entirely perturbative, and have behavior similar to TMD distributions with twist-two collinear distributions replaced by $1$. Alike $\Phi_{12}$ and $\Phi_{21}$ (\ref{small-b-tw-3}), the functions $\Psi_{12}$ and $\Psi_{21}$ behave as $\sim a_s b^\mu/b^2$ at $b\to 0$.

\subsection{Cancellation between IR and UV poles}
\label{sec:pole-cancel}

The renormalization of qTMD correlator (\ref{def:quasiTMD}) is
\begin{eqnarray}
\Omega^{[\Gamma]}_{\text{bare}}(\ell,b)=Z_W(b) Z_J^2\Omega^{[\Gamma]}(\ell,b,\mu).
\end{eqnarray}
The factor $Z_J$ is the renormalization of the heavy-to-light current (in the space-like regime). At NLO it reads \cite{Chetyrkin:2003vi}
\begin{eqnarray}\label{def:ZJ}
Z_J=1+\frac{a_s}{\epsilon}\frac{3}{2}C_F+\mathcal{O}(a_s^2).
\end{eqnarray}
The factor $Z_W$ trivially cancels between $\Psi$-functions and renormalization entirely. The cancellation of the remaining UV and IR divergences takes place individually for each current $\mathcal{J}$. So, one should have
\begin{eqnarray}
Z_J^{-1}C_1 Z^{\text{sub.}}_{U1}(\zeta)Z^{\text{sub.}}_\Psi(\bar \zeta)=\text{finite},
\qquad
Z_J^{-1}C_2 Z^{(0)\text{sub.}}_{U2}(\zeta)Z^{\text{sub.}}_\Psi(\bar \zeta)=\text{finite},
\end{eqnarray}
or
\begin{eqnarray}
&&\(\text{pole}[C_1]+Z^{\text{sub.}}_{U1}(\zeta)+Z^{\text{sub.}}_\Psi(\bar \zeta)-
Z_J\)_{\text{order }a_s}=0
\\
&&\(\text{pole}[C_2]+Z^{(0)\text{sub.}}_{U2}(\zeta)+Z^{\text{sub.}}_\Psi(\bar \zeta)-
Z_J\)_{\text{order }a_s}=0
\end{eqnarray}
where the last relations are valid only for $a_s$-order. The pole parts of coefficient functions are 
\begin{eqnarray}
\text{pole}[C_1]&=&\frac{a_s}{\epsilon}C_F\(-\frac{1}{\epsilon}-1-\ln\(\frac{\mu^2}{|2x(vP)|^2}\)+2\log\(i s s_x\)\),
\\
\text{pole}[C_2]&=&\frac{a_s}{\epsilon}\Big[
C_F\(-\frac{1}{\epsilon}+1-\ln\(\frac{\mu^2}{|2x(vP)|^2}\)+2\log\(i s s_x\)\)
\\\nn &&
-2\(C_F-\frac{C_A}{2}\)\frac{x_2+x_3}{x_2}\ln\(\frac{x_2+x_3-is0}{x_3-is0}\)\Big].
\end{eqnarray}
Now, using the NLO expressions for the renormalization constants (\ref{def:ZU1}, \ref{def:ZU2}, \ref{def:Zpsi}, \ref{def:ZJ}), we confirm the cancellation of poles if
\begin{eqnarray}\label{zeta*zeta}
\bar \zeta(\mu^2)\zeta =|2x(vP)|^2\mu^2.
\end{eqnarray}
This rule is universal for LP and NLP terms. The cancellation of IR and UV divergences for both the real and imaginary parts is a strong check of the computation of coefficient functions.

\subsection{Cancellation of special rapidity divergences and restoration of boost invariance}
\label{sec:boost-inv}

The special rapidity divergences cancel in the sum of term in the factorized expression. The cancellation is not traceless but leaves a term responsible for the restoration of boost-invariance of the whole expression. This important mechanism is not yet discussed in the literature, and thus we present it here with extra details.

The special rapidity divergences cancel in-between genuine NLP terms and do not require any additional ``soft-factor'' contribution (note that such a soft factor should carry an index $\mu$, and thus be a NNLP). This can be seen already from the Dirac structures of genuine and kinematic terms, which are richer for the genuine terms. Due to it, the number of Lorenz-invariant components of genuine terms is larger that those of kinematic terms. The explicit decomposition can be found in sec.\ref{sec:practice}. Each independent Lorentz-invariant component has a special rapidity divergences, but only some of them have contribution of smaller-twist functions that could be accompanied by some soft-factor to cancel it.  Thus, the cancellation of special rapidity divergences between genuine terms is the only possible mechanism which would work for all polarization cases.

The cancellation involves terms from several lines in the factorization formula (\ref{qTMD:bare-position}). To make it more explicit, we extract the   terms of interest here. Let us isolate the terms traced with $\llbracket \gamma^\mu\gamma^+\Gamma\rrbracket$, since the reasoning for the other combination is identical. We omit the superscript $\llbracket \gamma^\mu\gamma^+\Gamma\rrbracket$, the arguments $(\mu,\bar \zeta)$ for $\Psi$-functions, and $(\mu,\zeta)$ for $\Phi$-functions for better legibility. From \eqref{qTMD:bare-position} we have
\begin{eqnarray}\label{A1}
*=\Psi(b)i\partial_\mu \Phi_{11}(x,b)+i\Psi(b)\int \frac{[dx]}{x_2-is0}\delta(x-x_3)\Phi_{\mu,21}(x_{1,2,3},b)+\Psi^{(0)}_{\mu,21}(b)\Phi_{11}(x,b),
\end{eqnarray}
where all functions are renomalized.  The second and the third terms have special rapidity divergences. Adding and subtracting divergent terms, we promote distributions to their ``physical'' versions (\ref{TMD:phys}, \ref{Psi:phys})
\begin{eqnarray}\label{A2}
*&=&\Psi(b)i\partial_\mu \Phi_{11}(x,b)+i\Psi(b)\int \frac{[dx]}{x_2-is0}\delta(x-x_3)\mathbf{\Phi}_{\mu,21}(x_{1,2,3},b)+\mathbf{\Psi}^{(0)}_{\mu,21}(b)\Phi_{11}(x,b)
\\\nn &&
\qquad\qquad
- i[\partial_\mu \mathcal{D}(b)]\Psi(b)\Phi_{11}(x,b)\Big[\ln\(\frac{\delta^+}{q^+}\)-\ln\(\frac{\delta^-}{q^-}\)\Big],
\end{eqnarray}
where the last line contains the divergent terms. The rapidity renormalization parameters $\delta^\pm=\delta \nu^\pm$ and $q^\pm$ are not independent. The relation between them is fixed by the boost-invariance of the soft factor \cite{Vladimirov:2017ksc,Echevarria:2012js}. In terms of the boost-invariant combination of variables (see last line of (\ref{A2})), the relation reads
$$
\frac{\delta^+}{\delta^-}\frac{q^-}{q^+}=\sqrt{\frac{\bar{\zeta}}{\zeta}}
$$
In this way, we can rewrite the combination (\ref{A2}) as
\begin{eqnarray}\label{A3}
*&=&i\Psi(b)\(\partial_\mu-\frac{1}{2}[\partial_\mu\mathcal{D}(b)]\ln\(\frac{\bar \zeta}{\zeta}\)\) \Phi_{11}(x,b)
\\\nn &&+i\Psi(b)\int \frac{[dx]}{x_2-is0}\delta(x-x_3)\mathbf{\Phi}_{\mu,21}(x_{1,2,3},b)+\mathbf{\Psi}^{(0)}_{\mu,21}(b)\Phi_{11}(x,b).
\end{eqnarray}
This expression is written in the terms of boost-invariant $\zeta$ and $\bar \zeta$, and therefore, is independent on the used regulator for rapidity divergences. Each term in eqn. (\ref{A3}) is well-defined.

The combination that appears in the first line of (\ref{A3}) is not accidental. It is the only combination that supports the rescaling invariance for $\zeta$ and $\bar \zeta$, which is the consequence of the boost invariance. The factorization theorem fixes only the product $\zeta \bar \zeta$ (\ref{zeta*zeta}) and, therefore, it has to be invariant under the rescaling
\begin{eqnarray}\label{zeta-rescale}
\zeta \to \frac{\zeta}{\alpha},\qquad \bar \zeta\to \alpha\bar \zeta,
\end{eqnarray}
for any $\alpha\neq 0$. The rescaling invariance (\ref{zeta-rescale}) is obvious for the regular terms in TMD factorization, such as the LP term, and the terms in the second line of (\ref{A3}). It is straightforward to see it by differentiating the product $\Psi(\bar \zeta/\alpha)\Phi(\alpha \zeta)$ over $\alpha$ and applying rapidity evolution equations (\ref{evol:zeta}, \ref{Psi:rap.evol.}). The first line of (\ref{A3}) under the transformation (\ref{zeta-rescale}) transforms as
\begin{eqnarray}\label{zeta-rescale-2line}
\Psi(\bar \zeta)\(\partial_\mu -\frac{1}{2}[\partial_\mu \mathcal{D}]\ln\(\frac{\bar \zeta}{\zeta}\)\)\Phi(\zeta)
\to
\Psi(\alpha\bar \zeta)\(\partial_\mu -\frac{1}{2}[\partial_\mu \mathcal{D}]\ln\(\frac{\alpha^2\bar \zeta}{\zeta}\)\)\Phi\(\frac{\zeta}{\alpha}\),
\end{eqnarray}
where we omit all unnecessary arguments and indices for simplicity. The right-hand-side of (\ref{zeta-rescale-2line}) is independent on $\alpha$, which can be checked by differentiation and subsequent application of the equations (\ref{evol:zeta}, \ref{Psi:rap.evol.}). 

Throughout the above discussion, we omitted the coefficient functions. The reason is that the special rapidity divergences start at $\mathcal{O}(a_s)$. Therefore, the consideration presented here is valid at NLO. The inclusion of NLO coefficient will require the computation of special rapidity divergences at $a_s^2$-order, which goes beyond the scope of this work. However, the same formalism must be valid at all perturbative orders at NLP, unless the factorization theorem is broken.

\subsection{qTMD correlator at NLP (intermediate form)}
\label{sec:intemidiate-form}

Applying successively the procedures described in the previous section, namely, \textit{(i)} dividing by the soft factor, \textit{(ii)} combining the renormalization factors with the IR divergences of coefficient functions, \textit{(iii)} subtracting the divergent parts of integrals into $\mathbf{\Psi}$ and $\mathbf{\Phi}$;  we obtain the following expression for the renormalized qTMD correlator in momentum-fraction space
\begin{eqnarray}\label{qTMD:momentum-complex}
\Omega^{[\Gamma]}(x,b,\mu)&=&
\Psi (b;\mu,\bar \zeta)\mathbb{C}_{11}
\Phi_{11}^{\llbracket\Gamma\rrbracket}(x,b;\mu,\zeta)
\\\nn && 
+\frac{i}{2x P_+} \mathbb{C}_{11}\Psi (b) 
\(\partial_\mu-\frac{1}{2}[\partial_\mu \mathcal{D}(b,\mu)]\ln\(\frac{\bar \zeta}{\zeta}\)\) \Phi_{11}^{\llbracket\gamma^\mu\gamma^+\Gamma+\Gamma \gamma^+\gamma^\mu\rrbracket}(x,b;\mu,\zeta)
\\\nn &&
+\frac{i}{2xP_+}\Psi(b;\mu,\bar \zeta)\int \frac{[dx]}{x_2-is0}\Big[
\delta(x-x_3)\mathbb{C}_{21}(x,x_2)\mathbf{\Phi}_{\mu,21}^{\llbracket\gamma^\mu\gamma^+\Gamma\rrbracket}(x_{1,2,3},b;\mu,\zeta)
\\&&\nn
\qquad\qquad
+
\delta(x+x_1)\mathbb{C}_{12}(x,x_2) \mathbf{\Phi}_{\mu,12}^{\llbracket\Gamma\gamma^+\gamma^\mu\rrbracket}(x_{1,2,3},b;\mu,\zeta)\Big]
\\\nn &&
+\frac{1}{2xP_+}\mathbb{C}_{11v}
\mathbf{\Psi}^{(0)}_{\mu,21}(b;\mu,\bar \zeta)\Phi_{11}^{\llbracket\gamma^\mu\gamma^+\Gamma+\Gamma\gamma^+\gamma^\mu\rrbracket}(x,b;\mu,\zeta)
+ \mathcal{O}(\lambda^2).
\end{eqnarray}
where we have restored all arguments, and removed all regulators. The coefficient functions are
\begin{eqnarray}\label{C11}
\mathbb{C}_{11}&=&1+a_sC_F\(-\mathbf{L}_p^2-2\mathbf{L}_p-4+\frac{\pi^2}{6}\)+\mathcal{O}(a_s^2),
\\\label{C21}
\mathbb{C}_{21}(x,x_2)&=&
1+a_s\Big[C_F\(-\mathbf{L}_p^2+\frac{\pi^2}{6}+2\pi i s_x s\)
\\\nn &&
+2\(C_F-\frac{C_A}{2}\)\frac{x}{x_2}\ln\(\frac{x+is0}{x+x_2+is0}\)\(
\mathbf{L}_p+\ln\(\frac{x+is0}{x+x_2+is0}\)+2+i\pi s_x s\)
\\\nn &&
+2C_A\frac{x}{x+x_2}\ln\(\frac{-x-is0}{x_2-is0}\)
\Big]+\mathcal{O}(a_s^2),
\end{eqnarray}
\begin{eqnarray}
\label{C12}
\mathbb{C}_{12}(x,x_2)&=&(\mathbb{C}_{21}(x,-x_2))^*,
\\\label{C11v}
\mathbb{C}_{11v}&=&\mathbb{C}_{11}+\mathcal{O}(a_s^2),
\end{eqnarray}
where
\begin{eqnarray}
\mathbf{L}_p=\ln\(\frac{\mu^2}{4|x(vP)|^2}\),\qquad s_x=\sign(x).
\end{eqnarray}
The coefficient function $\mathbb{C}_{11}$ has been computed in refs.\cite{Ebert:2018gzl, Ebert:2019okf, Vladimirov:2020ofp}.

Let us stress that we equipped the term $\sim \ln(\bar\zeta/\zeta)$ by the coefficient function $\mathbb{C}_{11}$. This is a conjecture that does not follow from our NLO computation. As it is discussed in sec.\ref{sec:boost-inv}, the subtraction terms become sensitive to coefficient function only at $\mathcal{O}(a_s^2)$. However, this conjecture is supported by the boost invariance (\ref{zeta-rescale}) since the expression (\ref{qTMD:momentum-complex}) is the only that supports \eqref{zeta-rescale} exactly at all perturbative orders.

\subsection{Complex terms and TMD distributions of definite parity}
\label{sec:complex}

The expression for the factorization theorem (\ref{sec:intemidiate-form}) is not yet ready for a practical application. Since qTMD are real-valued functions, we expect that all the complex terms can be simplified into some real combinations. The resolution of the complex structure of NLP factorization is a straightforward but tedious procedure. Both TMD distributions and coefficient functions have complex parts.

The TMD distributions $\mathbf{\Phi}_{\mu,12}$ and $\mathbf{\Phi}_{\mu,21}$ are complex-valued functions with indefinite T-parity. A better choice of basis was suggested in ref.\cite{Rodini:2022wki}:
\begin{eqnarray}\label{def:definite-parity}
\mathbf{\Phi}^{[\Gamma]}_{\mu,\oplus}(x_1,x_2,x_3,b;\mu,\zeta)&=&
\frac{\mathbf{\Phi}^{[\Gamma]}_{\mu,21}(x_1,x_2,x_3,b;\mu,\zeta)
+\mathbf{\Phi}^{[\Gamma]}_{\mu,12}(-x_3,-x_2,-x_1,b;\mu,\zeta)}{2},
\\\nn
\mathbf{\Phi}^{[\Gamma]}_{\mu,\ominus}(x_1,x_2,x_3,b;\mu,\zeta)&=&
i\frac{\mathbf{\Phi}^{[\Gamma]}_{\mu,21}(x_1,x_2,x_3,b;\mu,\zeta)
-\mathbf{\Phi}^{[\Gamma]}_{\mu,12}(-x_3,-x_2,-x_1,b;\mu,\zeta)}{2}.
\end{eqnarray}
These functions have definite complexity and T-parity. For this reason they are called TMD distributions with definite parity. The drawback is that such functions do not have partonic interpretation, and mix during the evolution. Nonetheless, the basis $\{\mathbf{\Phi}_{\mu,\oplus}, \mathbf{\Phi}_{\mu,\ominus}\}$ is advantageous in comparison to $\{\mathbf{\Phi}_{\mu,12}, \mathbf{\Phi}_{\mu,21}\}$.

The complex part of the coefficient functions comes from differences sources, listed below.
\begin{itemize}
\item The terms $\sim i\pi s_x$ in the coefficient function $\mathbb{C}_{21}$.
\item The complex-valued logarithms in the coefficient function $\mathbb{C}_{21}$. Here and everywhere, we use the convention that the logarithm has a branch cut for the real negative argument. For example
\begin{eqnarray}
\ln(x-is0)=\ln|x|-is\pi\theta(-x),
\end{eqnarray}
where $\theta(x)$ is the Heaviside function.
\item The integration in the vicinity of $x_2=0$ point. It can be resolved by means of the ``plus''-distribution
\begin{eqnarray}
\frac{1}{x_2\pm is0}&=&\frac{1}{(x_2)_+}\mp i\pi s \delta(x_2),
\\
\frac{\ln(x_2\pm i0s)}{x_2\pm is0}&=&\(\frac{\ln|x_2|}{x_2}\)_++\frac{\pi^2}{2}\delta(x_2)\pm i\pi s \frac{\theta(-x_2)}{(x_2)_+},
\end{eqnarray}
where the ``plus''-distribution is defined as
\begin{eqnarray}
\int_{-\infty}^\infty dx f(x)(g(x))_+= \int_{-\infty}^\infty dx (f(x)-f(0))g(x).
\end{eqnarray}
\end{itemize}
Using these rules, and definition (\ref{def:definite-parity}), we rewrite the factorized expression (\ref{qTMD:momentum-complex}) in the explicitly real form. 

\subsection{QTMD correlator at NLP (final form)}
\label{sec:fac-final}

The final expression for the factorization of the qTMD correlator is
\begin{eqnarray}\label{qTMD:final}
\Omega^{[\Gamma]}(x,b,\mu)&=&
\Psi (b;\mu,\bar \zeta)\mathbb{C}_{11}
\Phi_{11}^{\llbracket\Gamma\rrbracket}(x,b;\mu,\zeta)
\\\nn && 
+\frac{i}{2x P_+} \mathbb{C}_{11}\Psi (b) 
\(\partial_\mu-\frac{1}{2}[\partial_\mu \mathcal{D}(b,\mu)]\ln\(\frac{\bar \zeta}{\zeta}\)\) \Phi_{11}^{\llbracket\gamma^\mu\gamma^+\Gamma+\Gamma \gamma^+\gamma^\mu\rrbracket}(x,b;\mu,\zeta)
\\\nn &&
+\frac{1}{2xP_+}\mathbb{C}_{11v}
\mathbf{\Psi}^{(0)}_{\mu,21}(b;\mu,\bar \zeta)\Phi_{11}^{\llbracket\gamma^\mu\gamma^+\Gamma+\Gamma\gamma^+\gamma^\mu\rrbracket}(x,b;\mu,\zeta)
\\\nn &&
+\frac{i}{2xP_+}\Psi(b;\mu,\bar \zeta)\int_{-1}^1 dx_2\Big[
\mathbb{C}_{R}(x,x_2)\mathbf{\Phi}_{\mu,\oplus}^{\llbracket\gamma^\mu\gamma^+\Gamma-\Gamma\gamma^+\gamma^\mu\rrbracket}(\tilde x,b;\mu,\zeta)
\\\nn  &&
\qquad\qquad\qquad\qquad\qquad\quad
+
s\pi\mathbb{C}_{I}(x,x_2)\mathbf{\Phi}_{\mu,\ominus}^{\llbracket\gamma^\mu\gamma^+\Gamma-\Gamma\gamma^+\gamma^\mu\rrbracket}(\tilde x,b;\mu,\zeta)
\\\nn  &&
\qquad\qquad\qquad\qquad\qquad\quad
-i \mathbb{C}_{R}(x,x_2)\mathbf{\Phi}_{\mu,\ominus}^{\llbracket\gamma^\mu\gamma^+\Gamma+\Gamma\gamma^+\gamma^\mu\rrbracket}(\tilde x,b;\mu,\zeta)
\\\nn  &&
\qquad\qquad\qquad\qquad\qquad\quad
+is\pi
\mathbb{C}_{I}(x,x_2)\mathbf{\Phi}_{\mu,\oplus}^{\llbracket\gamma^\mu\gamma^+\Gamma+\Gamma\gamma^+\gamma^\mu\rrbracket}(\tilde x,b;\mu,\zeta)
\Big]
\\\nn && + \mathcal{O}(\lambda^2),
\end{eqnarray}
where the integral over $x_2$ is restricted by the arguments of TMD distributions as:
$-1<x_2<1-x$ for $x>0$ and $-1-x<x<1$ for $x<0$. The argument of twist-three TMD distributions is $\tilde x=(-x-x_2,x_2,x)$ The coefficient functions are
\begin{eqnarray}\label{C11:final}
\mathbb{C}_{11}&=&1+a_sC_F\(-\mathbf{L}_p^2-2\mathbf{L}_p-4+\frac{\pi^2}{6}\)+\mathcal{O}(a_s^2),
\\\label{CR:final}
\mathbb{C}_{R}(x,x_2)
&=&\frac{1}{(x_2)_+}+a_s\Bigg\{
C_F\frac{-\mathbf{L}_p^2+\frac{\pi^2}{6}}{(x_2)_+}
\\\nn &&
+2\(C_F-\frac{C_A}{2}\)\frac{1}{(x_2)_+}\frac{x}{x_2}\Big[
\ln\(\frac{|x|}{|x+x_2|}\)\(\mathbf{L}_p+\ln\(\frac{|x|}{|x+x_2|}\)+2\)
\\\nn &&
-\pi^2(1+ s_x)\theta(x+x_2)-\pi^2(1-s_x)\theta(x)+2\pi^2\theta(x)\theta(x+x_2)\Big]
\\\nn &&
+2C_A\Big[
\frac{x}{x+x_2}\(\frac{\ln|x|}{(x_2)_+}-\(\frac{\ln|x_2|}{x_2}\)_+\)+\frac{\pi^2}{2}\delta(x_2)(2\theta(x)-s_x-1)\Big]\Bigg\}
+\mathcal{O}(a_s^2),
\\\label{CI:final}
\mathbb{C}_I(x,x_2)&=&\delta(x_2)+a_s\Bigg\{C_F\Big[
\delta(x_2)\(-\mathbf{L}_p^2+\frac{\pi^2}{6}\)
+\frac{2s_x}{(x_2)_+}\Big]
\\\nn &&+2\(C_F-\frac{C_A}{2}\)\Big[
\delta(x_2)\(-\mathbf{L}_p-2\)
+\frac{s_x}{(x_2)_+}\frac{x}{x_2}\ln\(\frac{|x|}{|x+x_2|}\)
\\\nn &&
-\frac{1}{(x_2)_+}\frac{x}{x_2}
\(\mathbf{L}_p+2\ln\(\frac{|x|}{|x+x_2|}\)+2\)(\theta(x)-\theta(x+x_2))\Big]
\\\nn &&
+2C_A\(\delta(x_2)\ln|x|-\frac{\theta(x)-\theta(-x_2)}{(x_2)_+}\frac{x}{x+x_2}\)\Bigg\}+\mathcal{O}(a_s^2).
\end{eqnarray}
The integrals with these coefficient functions are regular at all points of integration. 

The N$^2$LP correction, denoted by $\mathcal{O}(\lambda^2)$, includes the corrections N$^2$LP corrections in $P_+$ and NLP corrections in $1/L$. As we see here, the power corrections scale with $xP_+$ rather than just $P_+$. It agrees with our power counting (\ref{count:nbar}), defined for components of parton's momentum. Summarizing the factorization assumptions made in secs. \ref{sec:def} and \ref{sec:counting}, we specify
\begin{eqnarray}
\mathcal{O}(\lambda^2)=\mathcal{O}\(\frac{M^2}{x^2(vP)^2}, \frac{1}{b^2(vP)^2}, \frac{b}{L}, \frac{1}{ML}\).
\end{eqnarray}

The expression (\ref{qTMD:final}) is the complete NLP/NLO expression for the factorization of the qTMD correlator. It is apparently complicated and contains all possible combinations of factors and terms. Not all of these terms contribute to particular components of the qTMD correlator, as discussed in the next section. There are several interesting features of the expression (\ref{qTMD:final}) that are specific to the TMD factorization at NLP. 

The first feature is that NLP factorization mixes T-odd and T-even terms. The T-odd(even) TMD distributions (do not) change their global sign under the rotation of staple contour to a different-sign infinity \cite{Collins:2002kn}. The T-parity of $\Phi_\oplus^{[\Gamma]}$ is opposite to the T-parity of $\Phi_\ominus^{[\Gamma]}$. However, the terms with opposite parity in (\ref{qTMD:final}) always have relative factor $s$. Thanks to it, the relative sign between T-odd and T-even terms remains the same under the T-conjugation. Thus, NLP TMD factorization mixes distributions with different parity but preserves the global T-parity, which is required by the T-invariance of QCD.

The second feature is the presence of multiple $\theta$ functions. This is an unhealthy property of NLP TMD distributions. Their evolution equation also contains step functions, and due to it, the distributions are discontinuous at $x_i=0$. In particular, it leads to the appearance of the special rapidity divergences, discussed in sec.\ref{sec:boost-inv}. Nonetheless, the integrals are well-defined in (\ref{qTMD:final}). The structure of discontinuities could probably be simplified, but at the moment, such a procedure is unknown.

Finally, we observe that TMD factorization also incorporates the Qiu-Sterman-like contributions   \cite{Qiu:1991pp}, namely the contributions of twist-three distributions with the zero-momentum gluon $\mathbf{\Phi}(-x,0,x)$. Generally, twist-3 TMD distributions are discontinuous at $x_2=0$, but we have checked that, for all physically accessible cases of $\Omega^{[\Gamma]}$, the contributions  $\mathbf{\Phi}(-x,0,x)$ are well-defined for the known cases (see appendix C in ref.\cite{Rodini:2022wki}). The distributions $\mathbf{\Phi}$ are either continuous or zero at $x_2=0$, or contribute starting from $\mathcal{O}(a_s^2)$. 

The expression (\ref{qTMD:final}) is the first example of NLP TMD factorization at NLO written explicitly. The previous computations were either at LO \cite{Boer:2003cm,Balitsky:2017gis, Balitsky:2020jzt, Ebert:2021jhy}, or written in the abstract operator form \cite{Vladimirov:2021hdn}. It gives a taste of what one can expect from the NLP TMD factorization for other observables.

\section{On practical application of NLP factorization for qTMD correlators}
\label{sec:practice}

There are two main motivations to study the qTMD correlators. The first one is to determine the Collins-Soper kernel -- the nonperturbative function that governs the evolution in the rapidity scale of TMD distributions. The second one is to determine actual TMD distributions. In this section, we discuss the different possibilities of using NLP factorization theorem to improve our knowledge of TMD physics.

\subsection{Definition of qTMD distributions}
\label{sec:qTMD-distr}

Our starting point is the assumptions that the qTMD correlator can be computed on the lattice, as a function of $(vP)$, $\ell$ and $b^2$. Depending on $\Gamma$, it has different number of tensor components written in terms of the vectors $P^\mu$, $v^\mu$, $S^\mu$, $b^\mu$, and the tensors $g^{\mu\nu}$ and $\epsilon^{\mu\nu}_T$ \cite{Musch:2011er}. These components can be extracted individually. There are already several examples of such computations, see, f.i., \cite{Musch:2011er,Shanahan:2021tst,Schlemmer:2021aij,Shanahan:2020zxr,Engelhardt:2015xja}. The factorization theorem (\ref{qTMD:final}) provides the theoretical description for each component.

Generally speaking, the comparison can be made for any component of $\Omega^{[\Gamma]}$. However, some combination have a cleaner interpretation from the view-point of the factorization theorem. For example, the quasi-TMD correlators $\Omega^{[\gamma^0]}$ and $\Omega^{[\fnot v]}$ both give access to the unpolarized TMD distribution $f_1$. However, their sum is $\mathcal{O}(\lambda^2)$. Therefore, their difference has numerically smaller power-suppressed contribution, and is better suited to the study of leading nonperturbative physics. 

We introduce vectors $n^\mu$ and $\bar n^\mu$ in accordance to the definitions (\ref{def:n}, \ref{def:v}) (we set $v^2=-1$)
\begin{eqnarray}
\bar n^\mu=\(v^\mu+\frac{P^\mu}{(vP)}\)\frac{1}{\sqrt{2}\sqrt{1+\gamma^2}}-\frac{v^\mu}{\sqrt{2}}
\\
n^\mu=\(v^\mu+\frac{P^\mu}{(vP)}\)\frac{1}{\sqrt{2}\sqrt{1+\gamma^2}}+\frac{v^\mu}{\sqrt{2}},
\end{eqnarray}
where
$$
\gamma=\frac{M}{(vP)}.
$$
The convolutions with these vectors we denote as usual $a^+=(na)$ and $a^-=(\bar na)$ for any vector $a^\mu$. In addition, we define the symmetric and anti-symmetric transverse tensors
\begin{eqnarray}
g_T^{\mu\nu}&=&g^{\mu\nu}-\frac{v^\mu P^\nu+P^\mu v^\nu}{(vP)(1+\gamma^2)}+\frac{\gamma^2}{1+\gamma^2}\(v^\mu v^\nu-\frac{P^\mu P^\nu}{M^2}\),
\\
\epsilon_T^{\mu\nu}&=&-\frac{\epsilon^{\mu\nu\alpha\beta}v_\alpha P_\beta}{(vP)\sqrt{1+\gamma^2}}.
\end{eqnarray}
In this notation the vector of hadron's spin decomposes as
\begin{eqnarray}
S^\mu=\lambda \frac{P^\mu-\gamma M v^\mu}{M\sqrt{1+\gamma^2}}+S_T^\mu,\qquad \lambda=(vS)\gamma\sqrt{1+\gamma^2},
\end{eqnarray}
where $S_T$ is the transverse component $(S_TP)=(S_Tv)=0$. The main hard scale of the factorization is $P^+$, which in  terms of invariants reads
\begin{eqnarray}
P^+=\frac{(vP)}{\sqrt{2}}\(1+\sqrt{1+\gamma^2}\).
\end{eqnarray}
Note, that the difference between $(vP)$ and $P^+$ is $\sim \gamma^2$ and thus $\mathcal{O}(\lambda^2)$. Thus, without violation of counting one can use $(vP)$ instead of $P_+$.

Following refs.\cite{Ebert:2019okf, Ebert:2022fmh} we define qTMD distribution as
\begin{eqnarray}\label{def:qTMD-distr}
\widetilde{F}^{[\Gamma]}(\ell,b;\mu)&=&\frac{\widetilde{\Omega}^{[\Gamma]}(\ell,b;\mu)}{\Psi(b;\mu,\mu^2)}.
\end{eqnarray}
To distinguish a qTMD distribution from an ordinary TMD distribution we use the capital latter (instead of tilde as in ref.\cite{Ebert:2022fmh}, since the tilde-notation in this work is exclusively reserved to indicate the functions in position space). The transformation to the momentum-fraction space reads (\ref{def:Fourier-OMEGA})
\begin{eqnarray}
F^{[\Gamma]}(x,b;\mu)=\int_{-\infty}^{\infty} \frac{d\ell}{2\pi} 
e^{-ix\ell (vP)} \widetilde{F}^{[\Gamma]}(\ell,b,\mu)
=\frac{\Omega(x,b;\mu)}{\Psi(b,\mu,\mu^2)}.
\end{eqnarray}

The individual Dirac traces are parametrized as follows
\begin{eqnarray}\label{def:biq:V+}
F^{[\gamma^+]}&=&F_1+i\epsilon^{\mu\nu}_T b_\mu S_{T\nu}M F_{1T}^\perp,
\\\label{def:biq:A+}
F^{[\gamma^+\gamma^5]}&=&\lambda G_{1}+i(b \cdot S_T)M G_{1T},
\\\label{def:biq:T+}
F^{[i\sigma^{\alpha+}\gamma^5]}&=&S_T^\alpha H_{1}-i\lambda b^\alpha M H_{1L}^\perp
+i\epsilon^{\alpha\mu}b_\mu M H_1^\perp-\frac{M^2 b^2}{2}\(\frac{g_T^{\alpha\mu}}{2}-\frac{b^\alpha b^\mu}{b^2}\)S_{T\mu}H_{1T}^\perp,
\\\nn 
\label{def:biq:1}
F^{[\mathbbm{1}]}&=&\frac{M}{P^+}\Big[
E
+i\epsilon^{\mu\nu}_T b_\mu S_{T\nu} M\, E_T^\perp
\Big],
\\\label{def:biq:5}
F^{[i\gamma^5]}&=&\frac{M}{P^+}\Big[
\lambda E_L
+i(b\cdot S_{T}) M\, E_T
\Big],
\\\label{def:biq:V}
F^{[\gamma^\alpha]}&=&\frac{M}{P^+}\Big[
-\epsilon^{\alpha\mu}_TS_{T\mu} F_T
+i\lambda \epsilon^{\alpha\mu}b_\mu M\, F_L^\perp
-ib^\alpha M F^\perp
\\\nn &&
-b^2M^2\(\frac{g_T^{\alpha\mu}}{2}-\frac{b^\alpha b^\mu}{b^2}\)\epsilon_{T\mu\nu}S_T^\nu F_T^\perp
\Big],
\\\label{def:biq:A}
F^{[\gamma^\alpha\gamma^5]}&=&\frac{M}{P^+}\Big[
S_{T}^\alpha G_T
-i\lambda b^\alpha M\, G_L^\perp
+i\epsilon^{\alpha \mu}b_\mu M G^\perp
-b^2M^2\(\frac{g_T^{\alpha\mu}}{2}-\frac{b^\alpha b^\mu}{b^2}\)S_{T\nu} G_T^\perp
\Big],
\\\label{def:biq:T}
F^{[i\sigma^{\alpha\beta}\gamma^5]}&=&\frac{M}{P^+}\Big[
i(b^\alpha S_T^\beta-S_T^\alpha b^\beta)M H_T^\perp
-\epsilon^{\alpha\beta}_T H
\Big],
\\\label{def:biq:+-}
F^{[i\sigma^{+-}\gamma^5]}&=&\frac{M}{P^+}\Big[
\lambda H_L^\perp
+i(b\cdot S_T) M H_T
\Big],
\end{eqnarray}
where we omit the arguments $(x,b,\mu)$ of distributions on both sides. This parametrization is a straightforward generalization of the standard parametrization for ordinary TMD distributions \cite{Mulders:1995dh, Bacchetta:2006tn}. The remaining three Dirac traces ${F}^{[\gamma^-]}$, ${F}^{[\gamma^-\gamma^5]}$ and ${F}^{[i\sigma^{\alpha-}\gamma^5]}$ are $\mathcal{O}(\lambda^2)$ (parametrized by 8 distributions) and are not included in this list.

Comparing components (\ref{def:biq:V+} - \ref{def:biq:+-}) with the factorization theorem we find the factorization for each individual component. There are 8 qTMD distributions that obey the LP factorization, and 16 qTMD distributions that obey NLP factorization. 

\subsection{qTMD distributions with LP factorization}

The LP factorization theorem for qTMD distributions is well-understood and already applied in practice. The eight qTMD distributions that obey the LP factorization are those given in the lines (\ref{def:biq:V+}, \ref{def:biq:A+}, \ref{def:biq:T+}). This part of our computation coincides with the known results. For a review of the current state, see \cite{Ebert:2022fmh} and references within. In this subsection, we would like to provide a sketch of possible applications of the LP factorization theorem to contrast the problems with the application of the NLP factorization discussed in the following section.

The LP factorization theorem reads
\begin{eqnarray}\label{factorizationLP}
F(x,b;\mu)&=&\(\frac{(2|x|(vP))^2}{\zeta}\)^{-\mathcal{D}(b,\mu)}\mathbb{C}_{11}(\mathbf{L}_p,\mu)f(x,b;\mu,\zeta)+\mathcal{O}(\lambda^2),
\end{eqnarray}
where $F\in \{F_1, F_{1T}^\perp, G_1, G_{1T}, H_1, H_{1L}^\perp, H_1^\perp, H_{1T}^\perp\}$ and $f\in \{f_1, f_{1T}^\perp, g_1, g_{1T}, h_1, h_{1L}^\perp, h_1^\perp, h_{1T}^\perp\}$ in a natural one-to-one correspondence. The coefficient function $\mathbb{C}_{11}$ is given in eqn. (\ref{C11:final}). This factorization theorem has been derived in refs.\cite{Ebert:2019okf, Ji:2019ewn, Vladimirov:2020ofp, Ebert:2020gxr} using different techniques. In this work, we have explicitly demonstrated that the correction to (\ref{factorizationLP}) is $\mathcal{O}(\lambda^2)$, not $\mathcal{O}(\lambda)$.

The expression (\ref{factorizationLP}) is the simplest case among factorization formulas for qTMD distributions. The most direct application of (\ref{factorizationLP}) is the determination of Collins-Soper kernel from the ratio of qTMDs measured at different momenta $P$ \cite{Ebert:2019okf, Vladimirov:2020ofp}. One finds
\begin{eqnarray}
\frac{F(x,b;\mu;P_1)}{F(x,b;\mu;P_2)}=\(\frac{(vP_1)}{(vP_2)}\)^{-2\mathcal{D}(b,\mu)}\frac{\mathbb{C}_{11}(\mathbf{L}_{P1},\mu)}{\mathbb{C}_{11}(\mathbf{L}_{P2},\mu)}+\mathcal{O}(\lambda^2).
\end{eqnarray}
All ingredients of this expression, except $\mathcal{D}$, are perturbative, and thus Collins-Soper kernel can be determined. The precision of Collins-Soper kernel determined in this way is systematically improvable by increasing the perturbative order of $\mathbb{C}_{11}$ and the precision of lattice computation. This approach has been implemented in refs.\cite{Shanahan:2021tst, Shanahan:2020zxr}.

There is an alternative approach to determining the Collins-Soper kernel  \cite{Vladimirov:2020ofp}, which is technically much simpler but has limited precision. In this alternative approach, one considers the ratio of qTMDs directly in the position space. Limiting ourself to the case $\ell=0$, we find
\begin{eqnarray}\label{LP:ratio}
\frac{\widetilde{F}(\ell=0,b;\mu;P_1)}{\widetilde{F}(\ell=0,b;\mu;P_2)}=\(\frac{(vP_2)}{(vP_1)}\)^{2\mathcal{D}(b,\mu)}\mathbf{r}^{(0)}+\mathcal{O}(\lambda^2),
\end{eqnarray}
where
\begin{eqnarray}
\mathbf{r}^{(0)}=1+4C_F a_s
\ln\(\frac{(vP_1)}{(vP_2)}\)\[\ln\(\frac{\mu^2}{4(vP_1)(vP_2)}\)+1-2\mathbf{M}^{(0)f}_{\ln|x|}(b,\mu)\].
\end{eqnarray}
The function $\mathbf{M}$ is the ratio of integrals of TMD distributions
\begin{eqnarray}\label{LP:M}
\mathbf{M}^{(0)f}_{\ln|x|}(b,\mu)=\frac{\int_{-1}^1 dx \ln|x|\,|x|^{-2\mathcal{D}(b,\mu)}f(x,b;\mu,\zeta_0)}{\int_{-1}^1 dx |x|^{-2\mathcal{D}(b,\mu)}f(x,b;\mu,\zeta_0)},
\end{eqnarray}
where $f$ is the TMD distribution analogous to qTMD distribution $F$ (e.g. $f_1$ corresponds to $F_1$). The expression (\ref{LP:M}) is independent on $\zeta_0$.  In ref.\cite{Vladimirov:2020ofp}, it is argued that the nonperturbative function $\mathbf{M}$ is almost a constant in a broad range of $b$. This conjecture is supported by known phenomenological extractions. Therefore, the ``constant'' $\mathbf{M}$ can be fixed by comparing one of the lattice points (at $b\lesssim 1$GeV) to the perturbative value of $\mathcal{D}$. The method can be generalized to non-zero $\ell$. The detailed discussion can be found ref.\cite{Vladimirov:2020ofp}.

In this way, one avoids the decrease of precision due to the discrete Fourier transform over the lattice data and needs only a single $\ell$-value measurement. For the same reason, the method is technically much simpler. However, it contains an assumption $\mathbf{M}=\const$ with an unknown state, and its precision could not be improved beyond NLO (it requires an introduction of another unknown function analogous to $\mathbf{M}$). Nonetheless, the current systematic uncertainty of lattice simulations and the size of $\lambda$ are significant, and this method can be safely and reliably applied. It has been used in ref.\cite{Schlemmer:2021aij}.

The $\Psi$-function can be computed independently \cite{Ji:2019sxk, Li:2021wvl}. In this case, the factorization formula (\ref{factorizationLP}) can be used to determine the TMD distribution itself. For a more extended discussion of applications, we refer to recent reviews \cite{Ebert:2022fmh, Constantinou:2020hdm}. Note that the $\Psi$-function can be used as an independent source for the determination of Collins-Soper kernel \cite{LatticeParton:2020uhz, LPC:2022ibr}.

\subsection{qTMD distributions with NLP factorization}
\label{sec:practice-at-NLP}

The remaining 16 components of the qTMD correlator (\ref{def:biq:1}-\ref{def:biq:+-}) obey the NLP factorization. We write it in the following general form
\begin{eqnarray}\label{factorizationNLP}
F(x,b;\mu)&=&\frac{1}{x}\(\frac{(2|x|(vP))^2}{\zeta}\)^{-\mathcal{D}(b,\mu)}\Big\{
\mathbb{C}_{11}(\mathbf{L}_{p},\mu)A(x,b;\mu,\zeta)
\\\nn && +
\mathbb{C}_{11}(\mathbf{L}_{p},\mu)\Bigg(\mathbf{\Psi}_2(b)+\mathring{\mathcal{D}}(b)\ln\(\frac{\mu(2|x|(vP))}{\zeta}\)\Bigg)B(x,b;\mu,\zeta)
\\\nn &&
+\int_{-1}^1 dx_2\(\mathbb{C}_R(\mathbf{L}_p,x,x_2)C(\tilde x,b;\mu,\zeta)+s\pi\mathbb{C}_I(\mathbf{L}_p,x,x_2)D(\tilde x,b;\mu,\zeta)\)\Big\}
\end{eqnarray}
where $\tilde x=(-x-x_2,x_2,x)$. The letters $A$, $B$, $C$ and $D$ denote combinations of physical TMD distributions. They are listed in the table \ref{tab:tab} for each of NLP structure function. In general, $A$ and $B$ contain only twist-two distributions, and $C$ and $D$ contain only twist-three distributions.
We introduced also
\begin{eqnarray}
\mathbf{\Psi}_2(b)=\frac{ib^\mu}{b^2M^2}\frac{\mathbf{\Psi}_{\mu,21}^{(0)}(b;\mu,\mu^2)}{\Psi(b;\mu,\mu^2)},
\end{eqnarray}
which is dimensionless and scale-invariant (at least at NLO (\ref{def:AD:PSI})). The notation $\mathring{f}$ stands for 
\begin{eqnarray}
\mathring{f}(b)=\frac{2}{M^2}\frac{\partial f(b,\mu)}{\partial b^2},
\end{eqnarray}
for an arbitrary function $f$.
Note, that $\mathring{\mathcal{D}}(b)$ is independent on $\mu$ as a consequence of eqn. (\ref{integrability}). We stress that, in this representation, the factorization theorem is explicitly real-valued.

\begin{table}[t]
\renewcommand{\arraystretch}{2.}
\begin{center}
\resizebox{\textwidth}{!}{
\begin{tabu}{|c|c|[2pt]c|c|c|c|[2pt]c|}
\hline
$\Gamma$ & qTMD & $A$ & $B$ & $C$ & $D$ & T-odd
\\
\tabucline[2pt]{-}
\multirow{2}{*}{$\mathbbm{1}$} & $E$ &  & & $2\mathbf{h}_\oplus$ & $2\mathbf{h}_\ominus$ 
&
\\
\cline{2-7}
& $E_T^\perp$ &  & & $2\mathbf{h}_{\oplus T}^{A\perp}$ & $2\mathbf{h}_{\ominus T}^{A\perp}$
& \checkmark
\\
\tabucline[1.2pt]{-}
\multirow{2}{*}{$i\gamma^5$} &
$E_L$ &  & & $2\mathbf{h}_{\oplus L}^{}$ & $2\mathbf{h}_{\ominus L}^{}$& \checkmark
\\ 
\cline{2-7}
& $E_T$ &  & & $2\mathbf{h}_{\oplus T}^{D\perp}$ & $2\mathbf{h}_{\ominus T}^{D\perp}$ & \checkmark
\\
\tabucline[2pt]{-}
\multirow{4}{*}{$\gamma^\alpha$} &
$F_T$ &
$\Ds -f_{1T}^\perp-\frac{b^2M^2}{2}\mathring{f}_{1T}^\perp$  &
$\Ds\frac{b^2M^2}{2}f_{1T}^\perp$ & 
$\mathbf{f}_{\ominus T}^{}-\mathbf{g}_{\oplus T}^{}$ & 
$-\mathbf{f}_{\oplus T}^{}-\mathbf{g}_{\ominus T}^{}$
& \checkmark
\\
\cline{2-7}
& $F_L^\perp$ &  & & 
$-\mathbf{f}_{\ominus L}^{\perp}+\mathbf{g}_{\oplus L}^{\perp}$ & 
$\mathbf{f}_{\oplus L}^{\perp}+\mathbf{g}_{\ominus L}^{\perp}$
& \checkmark
\\
\cline{2-7}
& $F^\perp$ & 
$\mathring{f}_1$ & 
$-f_1$ & 
$\mathbf{f}_{\ominus}^{\perp}-\mathbf{g}_{\oplus}^{\perp}$ & 
$-\mathbf{f}_{\oplus}^{\perp}-\mathbf{g}_{\ominus}^{\perp}$
&
\\
\cline{2-7}
& $F_T^\perp$ & 
$\mathring{f}_{1T}^\perp$ & 
$-f_{1T}^\perp$ & 
$-\mathbf{f}_{\ominus T}^{\perp}+\mathbf{g}_{\oplus T}^{\perp}$ & 
$\mathbf{f}_{\oplus T}^{\perp}+\mathbf{g}_{\ominus T}^{\perp}$
& \checkmark
\\
\tabucline[1.2pt]{-}
\multirow{4}{*}{$\gamma^\alpha\gamma^5$} &
$G_T$ & 
$\Ds g_{1T}+\frac{b^2M^2}{2}\mathring{g}_{1T}$ &
$\Ds -\frac{b^2M^2}{2}g_{1T}$ & 
$-\mathbf{f}_{\oplus T}^{}-\mathbf{g}_{\ominus T}$ & 
$-\mathbf{f}_{\ominus T}^{}+\mathbf{g}_{\oplus T}^{}$
&
\\
\cline{2-7}
&$G_L^\perp$ & 
$\mathring{g}_1$ & 
$-g_1 $ & 
$\mathbf{f}_{\oplus L}^{\perp}+\mathbf{g}_{\ominus L}^{\perp}$ & 
$\mathbf{f}_{\ominus L}^{\perp}-\mathbf{g}_{\oplus L}^{\perp}$
&
\\
\cline{2-7}
&$G^\perp$ &  & & 
$\mathbf{f}_{\oplus}^{\perp}+\mathbf{g}_{\ominus}^{\perp}$ & 
$\mathbf{f}_{\ominus}^{\perp}-\mathbf{g}_{\oplus}^{\perp}$
& \checkmark
\\
\cline{2-7}
&$G_T^\perp$ &  
$\mathring{g}_{1T}$& 
$-g_{1T}$& 
$\mathbf{f}_{\oplus T}^{\perp}+\mathbf{g}_{\ominus T}^{\perp}$ & 
$\mathbf{f}_{\ominus T}^{\perp}-\mathbf{g}_{\oplus T}^{\perp}$
&
\\
\tabucline[1.2pt]{-}
\multirow{2}{*}{$i\sigma^{\alpha\beta}\gamma^5$} &
$H_T^\perp$ & 
$\Ds -h_{1T}^\perp+\mathring{h}_1-\frac{b^2M^2}{4}\mathring{h}_{1T}^\perp$ & 
$\Ds -h_1+\frac{b^2M^2}{4}h_{1T}^\perp$& 
$2\mathbf{h}_{\ominus T}^{A\perp}$ & 
$-2\mathbf{h}_{\oplus T}^{A\perp}$
&
\\
\cline{2-7} 
& $H$ & 
$-2h_{1}^\perp$ & & 
$-2\mathbf{h}_{\ominus}^{}$ & 
$2\mathbf{h}_{\oplus}^{}$
& \checkmark
\\
\tabucline[1.2pt]{-}
\multirow{2}{*}{$i\sigma^{+-}\gamma^5$} &
$H_L^\perp$ & 
$\Ds -2h_{1L}^\perp- b^2M^2 \mathring{h}_{1L}^\perp$  & 
$\Ds b^2 M^2 h_{1L}^\perp$ & 
$-2\mathbf{h}_{\ominus L}^{}$ & 
$2\mathbf{h}_{\oplus L}^{}$
&
\\
\cline{2-7}
& $H_T$ & 
$\Ds -h_{1T}^\perp -\mathring{h}_1-\frac{b^2M^2}{4}\mathring{h}_{1T}^\perp$ & 
$\Ds h_1 +\frac{b^2M^2}{4}h_{1T}^\perp $ & 
$-2\mathbf{h}_{\ominus T}^{D\perp}$ & 
$2\mathbf{h}_{\oplus T}^{D\perp}$
&
\\\hline
\end{tabu}
}
\caption{\label{tab:tab} The elements of the factorization theorem (\ref{factorizationNLP}). The empty cell corresponds to a vanishing element. The ``T-odd''-column indicates the qTMD distributions that change sign under $s\to-s$. The definitions of all distributions is given in appendix  \ref{app:parametrization}.}
\end{center}
\end{table}

The combinations of the TMD distributions that are present in $C$ and $D$ are not random. These combinations form autonomous pairs that mixes through the evolution \cite{Rodini:2022wki}. The evolution equations for TMD distributions of twist-three have an integral-differential form similar to the evolution of ordinary parton distributions but with an additional double-logarithmic term. The full set of TMD distributions of twist-three splits into subsets that evolve with the integral kernel $\mathbb{P}_A$ or $\mathbb{P}_B$ (see sec.4.3 in ref.\cite{Rodini:2022wki}). All combinations present in eqn. (\ref{factorizationNLP}) evolve with $\mathbb{P}_A$ only.

In comparison to the LP factorization formula (\ref{factorizationLP}) the NLP factorization formula (\ref{factorizationNLP}) is awkward. First, none of the qTMD distributions is proportional to a single TMD distribution of twist-three but always to a pair. Both coefficient functions $\mathbb{C}_R$ and $\mathbb{C}_I$ have tree-order contribution, and thus there is no perturbative suppression for one element of a pair. Second, many of the qTMD distributions contain twist-two terms, which could not be easily removed. Third, the function $\mathbf{\Psi}_2$ is a new nonperturbative component that cannot be determined solely from measurements of qTMD distributions. Thus, a direct determination of TMD distributions of twist-three from eqn.(\ref{factorizationNLP}) is cumbersome. 

Inspecting the table \ref{tab:tab}, we observe that each qTMD correlator has a counterpart with the same twist-three content. Therefore, by combining several qTMD correlators, one could disentangle individual components and determine the TMD distribution of twist-three. Here one should also account for the contamination by the twist-two terms. We found the following groups that share the same nonperturbative content
\begin{eqnarray}
&&\{E, H; H_1\},\qquad 
\{H_T^\perp, E_T^\perp; H_1, H_{1T}^\perp\},\qquad
\{H_L^\perp, E_L; H_{1L}^\perp\},\qquad \{H_T, E_T; H_1, H_{1T}^\perp \},
\\\nn &&
\{G_T, F_T, G_{1T}, F_{1T}^\perp\},\qquad
\{G_L^\perp, F_L^\perp; G_1\},\qquad
\{F^\perp, G^\perp; F_1\},\qquad
\{G_{T}^\perp, F_T^\perp; G_{1T}, F_{1T}^\perp\}.
\end{eqnarray}
In these sets, the first and the second elements are the T-even and T-odd qTMD distributions of sub-leading power correspondingly, and the last elements are LP qTMD distributions. However, even these combinations could not provide an unambiguous determination of twist-three distributions because the factorization formula projects twist-three functions to a single variable $x$.

A more immediate application can be made in the spirit of ref.\cite{Schlemmer:2021aij}, which is briefly explained below eqn.(\ref{LP:ratio}). Let us consider the ratio of in of distributions in the position space representation. For the cases with $B=0$, one shows that
\begin{eqnarray}\label{NLP:ratio}
\frac{\widetilde{F}(\ell=0,b;\mu;P_1)}{\widetilde{F}(\ell=0,b;\mu;P_2)}=\(\frac{(vP_2)}{(vP_1)}\)^{2\mathcal{D}(b,\mu)}\mathbf{r}_{\text{NLP}}^{(0)}(b,\mu)+\mathcal{O}(\lambda^2),
\end{eqnarray}
where
\begin{eqnarray}
\mathbf{r}_{\text{NLP}}^{(0)}(b,\mu)&=&1+4a_s(\mu)C_F\ln\(\frac{(vP_1)}{(vP_2)}\)\Big\{
 \ln\(\frac{\mu^2}{4(vP_1)(vP_2)}\)-2\mathbf{M}_{\text{NLP}}^{(0)F}(b,\mu)\Big\}.
\end{eqnarray}
The expression for the function $\mathbf{M}_{\text{NLP}}$ is rather lengthy and not instructive, so we do not write it here. Important is that $\mathbf{M}_{\text{NLP}}$ does not depend on $(vP)$, and, therefore, can be considered as a universal function. If we assume that $\mathbf{M}_{\text{NLP}}=\const$ (similarly to the LP case), then one can determine the Collins-Soper kernel using one of the lattice points for the normalization. The approach can be easily generalized to $\ell\neq0$ case if needed.

We cannot provide any justification for the assumption $\mathbf{M}_{\text{NLP}}=\const$, and if $A\neq 0$, this assumption is most probably too crude. If $A=0$ (these are the cases $\{E, E_T^\perp, E_L, E_T, F_L^\perp, G^\perp\}$) the expression for $\mathbf{M}_{\text{NLP}}$, although being  still complicated, it significantly simplifies in the large-$N_c$ limit:
\begin{eqnarray}\label{NLP:M}
\mathbf{M}_{\text{NLP}}\simeq 
\frac{\int_{-1}^1 \frac{dx}{x} \ln|x|\, |x|^{-2\mathcal{D}(b,\mu)}\int_{-1}^1 dx_2\(\frac{C(\tilde x,b;\mu,\zeta_0)}{(x_2)_+}+\delta(x_2)D(\tilde x,b;\mu,\zeta_0)\)}{\int_{-1}^1 \frac{dx}{x} |x|^{-2\mathcal{D}(b,\mu)}\int_{-1}^1 dx_2\(\frac{C(\tilde x,b;\mu,\zeta_0)}{(x_2)_+}+\delta(x_2)D(\tilde x,b;\mu,\zeta_0)\)}+\mathcal{O}\(\frac{1}{N_c^2}\).
\end{eqnarray}
This function has the same structure as the LP expression (\ref{LP:M}). Therefore, if the $b$-dependence does not significantly change as a function of $x$, one expects $\mathbf{M}_{\text{NLP}}\sim \const$. This assumption can be checked by comparing extractions of Collins-Soper kernels made from different pairs of $P_1$ and $P_2$.

Taking the same ratio (\ref{NLP:ratio}) in the momentum fraction space would only marginally simplify the ratio's structure. Importantly, the TMD distributions do not cancel entirely because the coefficient functions $\mathbb{C}_R$ and $\mathbb{C}_I$ depend on $x$ differently. Still, this difference is $\sim 1/N_c$, so we can write
\begin{eqnarray}\label{NLP:ratio2}
\frac{F(x,b;\mu;P_1)}{F(x,b;\mu;P_2)}&=&\(\frac{(vP_1)}{(vP_2)}\)^{2\mathcal{D}(b,\mu)}\Big[1
\\\nn &&
+4 a_s(\mu)C_F\ln\(\frac{(vP_1)}{(vP_2)}\)\ln\(\frac{\mu^2}{4|x|^2(vP_1)(vP_2)}\)
+\mathcal{O}\(\frac{a_s}{N_c}\)+\mathcal{O}\(a_s^2\)
\Big].
\end{eqnarray}
We remind that this formula is valid only if $A=0$ and $B=0$.

Concluding, the direct application of NLP factorization theorem (\ref{factorizationLP}) does not seem practical for the moment, due to its involved content that entangles several TMD distributions in a single qTMD distributions. Nonetheless, the ratios of qTMD distributions $E, E_T^\perp, E_L, E_T, F_L^\perp, G^\perp$ can provide access to the Collins-Soper kernel, in a way similar to the LP case. Such ratios can be considered both in position (\ref{NLP:ratio}) and in momentum (\ref{NLP:ratio2}) spaces. In both cases, ratios are not pure functions of $\mathcal{D}$ but contain contamination from twist-three TMD distributions. However, this contamination is small $\sim a_s/N_c$. In both cases, one could not improve the precision of the approach systematically (contrary to the LP case). We are hoping that further progress in studies of twist-three TMD distributions will open opportunities to use (\ref{factorizationLP}) more precisely.

\section{Conclusions}

In this work, we study a particular class of lattice observables known as quasi-transverse momentum-dependent (qTMD) distributions. These are the diagonal matrix elements between hadron states of a quark-quark correlator whose Wilson line is staple-like and equal-time. At large hadron's momentum, the qTMD correlator can be factorized in terms of physical TMD distributions and some unknown TMD-like functions. The form of the factorization theorem crucially depends on the Dirac matrix $\Gamma$ that contracts quark-fields' spinor indices. In this work, we consider two cases: $\Gamma\in\Gamma_+$ that projects both quark fields to their good components, and $\Gamma\in\Gamma_T$ that projects a good and a bad components of the quark field. The first case $\Gamma\in\Gamma_+$ obeys the leading-power (LP) factorization theorem and has already been studied in several works. The case $\Gamma\in\Gamma_T$ requires the next-to-leading power (NLP) factorization and is addressed in this work for the first time.

We derive the factorization theorem for qTMDs with $\Gamma\in \Gamma_T$ and compute the corresponding coefficient functions at NLO. For the first time, we present the outcome of TMD factorization at NLP/NLO in a directly usable form. In this sense, the expression derived in this work can serve as an example of a structure expected for more involved observables, such as differential cross-sections. We explicitly demonstrate that by combining all elements of NLP TMD factorization, one obtains a valid and well-defined expression. It is not a trivial statement due to the different singularity structures between NLP and LP cases. The computation is done for a general $\Gamma\in\Gamma_T$, which includes 16 different qTMD distributions measurable on the lattice. As a by-product, we also obtain the LP factorization and confirm previous computations with a different method. We explicitly check that the NLP factorization theorem does not contribute to the case $\Gamma\in\Gamma_+$, and thus any power correction to them actually starts at N$^2$LP. Note, that in this work, we operate with massless quarks ignoring power corrections of the type $m_q/(vP)$.

Along the work, we made several observations related to NLP TMD factorization that are general and important beyond the physics of qTMD distributions. 
\begin{itemize}
\item We observed that, at the bare NLO level, the NLP coefficient functions exactly reproduce the LP coefficient function in the limit of vanishing gluon momentum. We have checked that the same observation holds for the bare coefficient functions in Drell-Yan/SIDIS \cite{Vladimirov:2021hdn}. Since at NLP $v$-collinear gluons carry vanishing light-cone momentum, the NLP coefficient function $C_v$ is equal to the LP coefficient $C_1$. We argue that these relations could be a consequence of soft-gluon theorems and valid at all perturbative orders, but we do not have general proof of this statement beyond NLO.
\item For the first time, we explicitly demonstrate the cancellation of special rapidity divergences. The special rapidity divergences appear in the integral convolutions of twist-three TMD distributions and were observed in ref.\cite{Rodini:2022wki}. The cancellation takes place in-between different collinear sectors and restores the boost invariance of the NLP factorization theorem. This mechanism is an essential part of the proof of the TMD factorization at NLP.
\item In the NLP factorization theorem, a non-trivial interplay occurs between the real and imaginary parts of the coefficient functions and the TMD parametrizations. Consequently, TMD factorization also incorporates Qiu-Sterman-like contributions, i.e., contributions of twist-three distributions with vanishing gluon momentum. Such contributions appear already at LO and were missed in many previous considerations.
\end{itemize}

The expression for the TMD factorization theorem at NLP is rather more complex than its LP counterpart. It mixes TMD distributions of twist-two (and their derivatives), twist-three, and derivatives of Collins-Soper kernels. Due to it, the direct application of the theorem is involved and requires several measurements to disentangle individual elements. However, even in this case, the determination will be incomplete because qTMD distribution depends on a single momentum fraction, whereas a twist-three distribution depends on two momentum fractions. Yet we demonstrate that a subset of observables (6 out of 16) can be used individually to determine the Collins-Soper kernel using a simplified procedure. The procedure is valid in the large-$N_c$ approximation and assumes that some integral weakly depends on the transverse distance. Both assumptions are accurate at the current precision of lattice simulations.

\acknowledgments
We thank Andreas Sch\"afer for numerous discussions and the motivation to study this case. A.V. is funded by the \textit{Atracci\'on de Talento Investigador} program of the Comunidad de Madrid (Spain) No. 2020-T1/TIC-20204. A.V. is also supported by the Spanish Ministry grant PID2019-106080GB-C21. This work was partially supported by DFG FOR 2926 ``Next Generation pQCD for  Hadron  Structure:  Preparing  for  the  EIC'',  project number 430824754. S.R. acknowledge the financial support from the physics department of Ecole Polytechnique.

\appendix

\section{Standard parametrization of TMD distributions}
\label{app:parametrization}
In this appendix we collect the parametrizations for the LP and NLP TMD correlators. These parametrizations have bees used to derive the table \ref{tab:tab}.
The spin vector is written as:
\begin{eqnarray}
s^\mu=\lambda\frac{P^- n^\mu-P^+ \bar n^\mu}{M}+s_T^\mu,
\end{eqnarray}
where $M$ is the mass of the hadron. It implies $\lambda=M s^+/p^+$.
The standard parameterization of the leading twist TMD correlators has been carried out in ref.~\cite{Mulders:1995dh}. We recall the parameterization here for completeness and consistency.
\begin{eqnarray}\label{def:TMDs:1:g+}
\Phi^{[\gamma^+]}(x,b)&=&f_1(x,b)+i\epsilon^{\mu\nu}_T b_\mu s_{T\nu}M f_{1T}^\perp(x,b),
\\\label{def:TMDs:1:g+5}
\Phi^{[\gamma^+\gamma^5]}(x,b)&=&\lambda g_{1}(x,b)+i(b \cdot s_T)M g_{1T}(x,b),
\\\label{def:TMDs:1:s+}
\Phi^{[i\sigma^{\alpha+}\gamma^5]}(x,b)&=&s_T^\alpha h_{1}(x,b)-i\lambda b^\alpha M h_{1L}^\perp(x,b)
\\\nn && +i\epsilon^{\alpha\mu}b_\mu M h_1^\perp(x,b)-\frac{M^2 b^2}{2}\(\frac{g_T^{\alpha\mu}}{2}-\frac{b^\alpha b^\mu}{b^2}\)s_{T\mu}h_{1T}^\perp(x,b),
\end{eqnarray}
where $b^2<0$. All TMD distributions are dimensionsless real functions that depend on $b^2$ (the argument $b$ is used for shortness).

At sub-leading power, we parametrize the correlators with definite T-parity given in Eq.~\eqref{def:definite-parity} as follow (see \cite{Rodini:2022wki})
\begin{eqnarray}\label{def:TMDs:2:g+}
\Phi_{\bullet}^{\mu[\gamma^+]}(x_{1,2,3},b)&=&
\epsilon^{\mu\nu}s_{T\nu} M f_{\bullet T}(x_{1,2,3},b)
+ ib^\mu M^2 f^\perp_\bullet(x_{1,2,3},b)
\\\nn &&
+i\lambda \epsilon^{\mu\nu}b_\nu M^2 f^\perp_{\bullet L}(x_{1,2,3},b)
+b^2M^3\epsilon_T^{\mu\nu}\(\frac{g_{T,\nu\rho}}{2}-\frac{b_\nu b_\rho}{b^2}\)s^{\rho}_Tf_{\bullet T}^\perp(x_{1,2,3},b),
\\\label{def:TMDs:2:g+5}
\Phi_{\bullet}^{\mu[\gamma^+\gamma^5]}(x_{1,2,3},b)&=&
s_T^\mu M g_{\bullet T}(x_{1,2,3},b)
-i\epsilon^{\mu\nu}_Tb_\nu M^2 g^\perp_\bullet(x_{1,2,3},b)
\\\nn &&
+i\lambda b^\mu M^2 g_{\bullet L}^\perp(x_{1,2,3},b)
+b^2M^3\(\frac{g_T^{\mu\nu}}{2}-\frac{b^\mu b^\nu}{b^2}\)s_{T\nu}g_{\bullet T}^\perp(x_{1,2,3},b),
\\\nn
\Phi_{\bullet}^{\mu[i\sigma^{\alpha+}\gamma^5]}(x_{1,2,3},b)&=&
\lambda g_{T}^{\mu\alpha} M h_{\bullet L}(x_{1,2,3},b) 
+\epsilon^{\mu\alpha}_T M h_\bullet(x_{1,2,3},b)
+ig_T^{\mu\alpha}(b\cdot s_T)M^2 h_{\bullet T}^{D\perp}(x_{1,2,3},b)
\\\nn &&
+i(b^\mu s^\alpha_T-s_T^\mu b^\alpha)M^2h_{\bullet T}^{A\perp}(x_{1,2,3},b)
+(b^\mu \epsilon^{\alpha\beta}_Tb_\beta+\epsilon_T^{\mu\beta}b_\beta b^\alpha)M^3h_{\bullet}^\perp(x_{1,2,3},b)
\\\label{def:TMDs:2:s+} &&
+\lambda M^3 b^2\(\frac{g^{\mu\alpha}_T}{2}-\frac{b^\mu b^\alpha}{b^2}\)h_{\bullet L}^\perp(x_{1,2,3},b)
\\\nn &&
+i(b\cdot s_T) M^2\(\frac{g^{\mu\alpha}_T}{2}-\frac{b^\mu b^\alpha}{b^2}\)h_{\bullet T}^{T\perp}(x_{1,2,3},b)
\\\nn &&
+iM^2\(\frac{b^\mu s^\alpha_T+s_T^\mu b^\alpha}{2}-\frac{b^\mu b^\alpha}{b^2}(b\cdot s_T)\)h_{\bullet T}^{S\perp}(x_{1,2,3},b)
,
\end{eqnarray}
The distributions defined in (\ref{def:TMDs:2:g+}, \ref{def:TMDs:2:g+5}, \ref{def:TMDs:2:s+}) are dimensionless and real functions.   
The notation for the TMD distributions follows the traditional pattern used in the parameterization of leading TMD distributions (\ref{def:TMDs:1:g+}, \ref{def:TMDs:1:g+5}, \ref{def:TMDs:1:s+}). Namely, the proportionality to $b$ is marked by the superscript $\perp$, and the polarization by subscript $L$ (for longitudinal) or $T$ (for transverse). In the tensor case, there are four structures $\sim b^\mu s^\alpha_T$, which are denoted as $h_T^{A\perp}$, $h_T^{D\perp}$, $h_T^{S\perp}$, $h_T^{T\perp}$ for antisymmetric, diagonal, symmetric, and traceless components. In total there are 32 TMD distributions of twist-three. Among the 32 TMD distributions, 16 distributions change the sign under T-parity transformation, and 16 do not. It means that 16 distributions are na\"ively T-odd. For a complete classification of the NLP power TMD we refer to ref. \cite{Rodini:2022wki}.
\bibliography{bibFILE}
\end{document}